\documentclass[11pt]{article}
\usepackage{amsbsy,amssymb,amsmath,bm,epsfig,rotating,wrapfig}

\renewcommand\baselinestretch{1.10}
\def\qprop{\textsc{Qprop}}

\def\imagi{\mathrm{i}}
\def\diff{\mathrm{d}}

\def\halb{\frac{1}{2}}

\def\pabl#1#2{\frac{\partial #1}{\partial #2}}

\def\energy{{\cal E}}

\def\bE{{\mathbf E}} 
\def\be{{\mathbf e}} 
\def\br{{\mathbf r}} 
\def\bp{{\mathbf p}} 
\def\bA{{\mathbf A}}

\def\rY{{\mathrm Y}}

\def\reff#1{(\ref{#1})}

\def\beq{\begin{equation}}
\def\eeq{\end{equation}}

\begin{document}

\title{\qprop: A Schr\"o{}dinger-solver for intense laser-atom interaction}

\author{Dieter Bauer\footnote{Email: dbauer@mpi-hd.mpg.de}\ ,  Peter Koval \\
 {\small\em   Max-Planck-Institut f\"ur Kernphysik, Postfach 103980, 69029 Heidelberg, Germany}}

\maketitle

\date{}

\setlength{\parindent}{0.5cm}

\begin{abstract}
The \qprop\ package is presented. \qprop\ has been developed to study laser-atom interaction in the nonperturbative regime where nonlinear phenomena such as above-threshold ionization, high order harmonic generation, and dynamic stabilization are known to occur. In the nonrelativistic regime and within the single active electron approximation, these phenomena can be studied with \qprop\ in the most rigorous way by solving the time-dependent Schr\"odinger equation in three spatial dimensions.
Because \textsc{Qprop} is optimized for the study of quantum systems that are spherically symmetric in their initial, unperturbed configuration, all wavefunctions are expanded in spherical harmonics.
Time-propagation of the wavefunctions is performed using  a split-operator approach. Photoelectron spectra are calculated employing a window-operator technique.
Besides the solution of the time-dependent Schr\"odinger equation in single active electron approximation, \qprop\ allows to study many-electron systems via the solution of the time-dependent Kohn-Sham equations. 
\end{abstract}

\newpage
\setlength{\parindent}{0.0cm}

\textbf{\large PROGRAM SUMMARY}

\bigskip

\textit{Title of program:} \textsc{Qprop}.

\bigskip

\textit{Catalogue number:} To be assigned.

\bigskip

\textit{Program obtainable from:} CPC Program Library, 
     Queen's University of Belfast, N. Ireland.
     
\bigskip

\textit{Computer on which program has been tested:} PC Pentium IV, Athlon.

\textit{Operating system:} Linux.
     
\bigskip
     
\textit{Program language used:} \textsc{C++}.

\bigskip

\textit{Memory required to execute with typical data:} Memory requirements 
depend on the number of propagated orbitals and on the size of the orbitals.
For instance, time-propagation of a hydrogenic wavefunction in the perturbative regime requires about 64 KByte RAM (4 radial orbitals with 1000 grid points).
Propagation in the strongly nonperturbative regime providing energy spectra up to high energies may need 60 radial orbitals, each with 30000 grid points, i.e., about 30 MByte. Examples are given in the article.
\bigskip

\textit{No.\ of bits in a word:} Real and complex valued numbers
of double precision are used.

\bigskip

\textit{Peripheral used:} Disk for input--output, terminal for 
interaction with the user.

\bigskip

\textit{CPU time required to execute test data:} Execution time depends
on the size of the propagated orbitals and the number of time-steps. 

\bigskip

\textit{Distribution format:} Compressed tar archive.

\bigskip


\bigskip

\textit{Nature of the physical problem:} Atoms put into the strong field 
of modern lasers display a wealth of novel phenomena that are not accessible to conventional perturbation theory where the external field is considered small as compared to inneratomic forces. Hence, the full {\em ab initio} solution of the time-dependent Schr\"odinger equation is desirable but in full dimensionality only feasible for no more than two (active) electrons. If many-electron effects come into play or effective ground state potentials are needed, (time-dependent) density functional theory may be employed. \textsc{Qprop} aims at providing tools for {\em (i)} the time-propagation of the wavefunction according to the time-dependent Schr\"odinger equation,  {\em (ii)} the time-propagation of Kohn-Sham orbitals according to the time-dependent Kohn-Sham equations, and {\em (iii)} the energy-analysis of the final one-electron wavefunction (or the Kohn-Sham orbitals).

\bigskip

\textit{Method of solution:}  
An expansion of the wavefunction in spherical harmonics leads to a coupled set of equations for the radial wavefunctions.
These radial wavefunctions are propagated using a split-operator technique and the Crank-Nicolson approximation for the short-time propagator.   The initial ground state is obtained
via imaginary time-propagation for spherically symmetric (but otherwise arbitrary) effective potentials. Excited states can be obtained through the combination of imaginary time-propagation and orthogonalization. For the 
Kohn-Sham scheme a multipole  expansion of the effective potential is employed. Wavefunctions can be analyzed using the window-operator technique, facilitating the calculation of electron spectra, either angular-resolved or integrated.

\bigskip 

\textit{Restrictions onto the complexity of the problem:}
The coupling of the atom to the external field is treated in dipole approximation. The time-dependent Schr\"odinger solver is restricted to the treatment of a single active electron.  As concerns the time-dependent density functional mode of \qprop,
the Hartree-potential (accounting for the classical electron-electron repulsion) is expanded up to the quadrupole. Only the monopole term of the Krieger-Li-Iafrate
exchange potential is currently implemented. As in any nontrivial optimization problem, convergence to the optimal many-electron state (i.e., the ground state) is not automatically guaranteed.

\bigskip
    
\textit{External routines/libraries used:} The program uses 
the well established  libraries \textsc{blas}, \textsc{lapack}, and \textsc{f2c}.

\bigskip

\setlength{\parindent}{0.5cm}

\section{\label{s:introduction}Introduction}

Progress in the construction of lasers brings powerful sources of electromagnetic radiation into the 
laboratory. The frequency domain of modern lasers covers a wide range from the infrared to the ultraviolet. Laser intensities of $10^{16}$ W/cm$^2$ and more are routinely achieved in many laboratories all over the world \cite{Agostini-and-DiMauro:2004}. Such powerful light causes
a number of strong-field phenomena like tunneling and above-threshold
ionization,  high order harmonic generation, nonsequential ionization, pronounced dynamic Stark-shifts, and the (not yet experimentally confirmed) dynamic stabilization \cite{Agostini-and-DiMauro:2004,becker:2002,Joachain-etal:2000,brabec:2000,Delone-Krainov:1999,proto:1997}. All these phenomena are not accessible through ``conventional'' perturbation theory where the external laser field is required to be small compared to the atomic binding force acting on the electron.

In the nonrelativistic regime (i.e., moderate charge states and moderate laser intensities $< 10^{18}$\,Wcm$^{-2}$), the solution of the time-dependent Schr\"odinger equation is the most rigorous nonperturbative approach to the problem of intense laser-atom interaction. Fortunately, most of the above mentioned phenomena (apart from nonsequential ionization) can be understood within a single active electron picture, making a numerical {\em ab initio} treatment possible at all. However, since an electron released by an intense laser may travel thousands of atomic units during the pulse, huge numerical grids in position space are needed. It was thus natural to investigate first low-dimensional model systems where the three electronic degrees of freedom were restricted to the laser polarization direction (see, e.g., \cite{Su-and-Eberli:1991,
Pen-and-Jiang:1992,Chen-and-Bernstein:1993,Grobe-and-Eberly:1993,Pindzola-etal:1995}).  Much qualitative insight was gained by these model studies while the quantitative agreement with experiment or full three-dimensional calculations is, in general, poor.   

With continuously increasing computer power, two- and
three-dimensional models of one and two-electron atoms were investigated. Two and three-dimensional studies of the single active electron dynamics in intense laser fields permit the use of elliptically polarized drivers, the calculation of angular distributions, and the investigation of the polarization properties of the light emitted into (high order) harmonics.
 
The methods used for the time-dependent propagation of the electronic wavefunction have been improved to lower the computational demand. Examples are variable time-
and space stepping \cite{Cerjan-and-Kosloff:1993,Tong-and-Chu:1997},
advanced finite difference approximations
\cite{Smyth-etal:1998,Muller:1999},
advanced interpolation methods like Gauss-Legendre or
\textit{B}-Spline interpolation \cite{Tong-and-Chu:1997,Bachau-etal:2001}, higher-order time propagators \cite{Puzynin-etal:2000,
Cerjan-and-Kulander:1991}, or matrix iterative methods \cite{nurhuda:1999}.

There exist alternative methods, which avoid the propagation of the time-dependent wave function on a numerical grid representing real position space.  Instead,  the $(t,t')$-method \cite{peskin}, Floquet theory \cite{potvliege,chu}, and complex scaling \cite{reinh,scrinzi,chu} make use of extended Hilbert spaces, (approximate) time-periodicity, or complex continuation of position space, respectively.

Codes propagating two-electron 
wavefunctions in their full dimensionality are currently at the limit of feasibility
\cite{Smyth-etal:1998,scrinzi,Meharg-etal:2005}. It is unlikely that more complex systems will be treated on an equal
footing because of the exploding computational demand. Even if the final, high-dimensional, multi-electron wavefunction were available, its analysis would pose new problems. In other words, the brute-force approach to multi-electron atoms in intense laser fields is a dead end    \cite{Kohn:1999}.

It is the Nobel-prize winning density-functional theory (DFT) that provides a genuine resort for quantum mechanics to treat many-electron systems as {\em ab initio} as possible. The central theorem of DFT, the Hohenberg-Kohn theorem, states that quantum mechanics may be based on the electron density alone.   The density remains a three-dimensional quantity, independently
of the number of particles in the system. As a consequence, density-functional theory does not suffer from the exponential explosion of computational demand, the so-called exponential wall \cite{Kohn:1999}.
In fact, DFT is now well-established in electronic structure calculations and quantum chemistry \cite{Kohn:1999,dreizler:1990,Parr:1989,Joubert:1998,Fiolhais-etal:2003}.  Most of the modern formulations of density-functional theory
rely on the Kohn-Sham equations which provide a practicable computational scheme. 
Density-functional theory has been extended to the time-dependent domain \cite{runge:1984,burke:1998,ullrich:1996,marques:2003}
and a number of algorithms for time-dependent Kohn-Sham-solvers have been published \cite{Marques-etal:2003, Burdick-etal:2003,Tong-and-Chu:1998}. 

Two of the most interesting observable quantities are the spectrum of the radiation emitted by the atom and photoelectron spectra. The former can be easily calculated via the Fourier-transformation of the dipole acceleration while the latter is much more tricky. In principle, one should project the final wavefunction onto the (continuum) eigenstates of the unperturbed Hamiltonian. This is numerically very demanding, especially if the continuum eigenstates are analytically unknown. In contrast to the straightforward projection, implicit methods to calculate the photoelectron spectra can be of distinct advantage. In the \textsc{Qprop} code, a window-operator technique \cite{Schafer-and-Kulander:1990,Schafer:1991}  for the energy-analysis of the final wavefunction is employed, facilitating the calculation of total and angular-resolved photoelectron spectra without explicit determination of eigenstates.

The aim of this
paper is to present a collection of routines that 
consistently realize three issues:
{\em (i)} time-propagation of a wavefunction according to the time-dependent Schr\"odinger equation,  {\em (ii)} time-propagation of Kohn-Sham orbitals according to the time-dependent Kohn-Sham equations, and {\em (iii)} the energy-analysis of the final one-electron wavefunction (or the Kohn-Sham orbitals).
The actual algorithm for time 
propagation has been taken from \cite{Muller:1999}, where it was introduced explicitly for the case of a linearly polarized laser field. 
In \textsc{Qprop}, we generalized the algorithm in two directions: {\em (i)} elliptic polarization, and {\em (ii)} effective many-electron potentials. 
For the exchange-part of the effective Kohn-Sham potential we use the expression suggested by Krieger, Li, and Iafrate \cite{Krieger-etal:1992}.
The implementation of the window-operator routine for the spectral analysis of wavefunctions was inspired by the work of Schafer and Kulander  \cite{Schafer-and-Kulander:1990}.

\qprop\ does not posses a fixed input/output interface but rather is a library, functions
of which realize the time-propagation and the analysis of the wavefunction.
Hence, in order to profit from the program, the user should be familiar with the basics of the underlying theory and the algorithm. Both are presented
in Secs.~\ref{s:theory} and \ref{s:program-structure}. Moreover,
the five examples in Sec.~\ref{s:test-calculations} should ease the first steps. Additional examples, patches, updates, and further documentation are provided elsewhere \cite{qpropwebpage}. The \textsc{Qprop} package has been successfully applied to a number of problems \cite{bauer-etal:2005,milo-etal:2005,Bauer-etal:2005,Lindner-etal:2005,Bauer:2005,Bauer-and-Ceccherini:2002,Bauer-and-Ceccherini:2001}.

\section{\label{s:theory}Theoretical background}

\subsection{\label{ss:1.1}Time-dependent first-principle equations}
The nonrelativistic quantum dynamics of an electron is governed by the time-dependent Schr\"odinger equation, i.e.,  the electron wavefunction $\Psi(\br t)$ obeys\footnote{Atomic units ($m_e=\hbar=\vert e\vert=1$) are used unless noted otherwise.}
\begin{equation}
\imagi\frac{\partial \Psi(\br t)}{\partial t}=H \Psi(\br t).
\label{TDSE}
\end{equation}
Here, spin degrees of freedom are neglected. The Hamiltonian $H$ 
may depend on time $t$ and in general
reads
\begin{equation}
H=H_{\mathrm{S}}=\left( \frac{1}{2}[\bp+\bA(\br t)]^2
+ \phi(\br t)\right)
\label{s-h}
\end{equation}
where $\bA(\br t)$ and $\phi(\br t)$ are vector and scalar potential, respectively.
The operator $\bp$ is the canonical momentum while $\bp'=\bp+\bA(\br t)$ is the kinetic momentum so that ${\bp'}^2/2=[\bp+\bA(\br t)]^2/2$ is the kinetic energy. 
Extending the Schr\"o{}dinger equation \reff{TDSE} to
$N$-electron systems is formally straightforward, leading to a $N$-electron wavefunction  in a $3N$-dimensional Hilbert space (spin neglected). In practice, however, current super computers are hardly capable of treating two-electron atoms in full dimensionality exposed to strong laser fields \cite{Meharg-etal:2005}, let alone systems with more than two electrons.

Time-dependent density-functional theory (TDDFT) offers to reduce the numerical effort dramatically \cite{runge:1984,burke:1998,ullrich:1996,marques:2003}.
The basic quantity of density-functional
theory is the electron density $n(\br t)$, which is always
a function of three space coordinates and time, independent of the number of particles $N$.
In practice, time-dependent
density functional theory is formulated via the
time-dependent Kohn-Sham equation
\begin{equation}
\imagi\frac{\partial \Psi_i(\br t)}{\partial t}=H_\mathrm{KS} \Psi_i(\br t), \qquad i=1,2, \ldots N,
\end{equation}
describing the evolution of the 
Kohn-Sham orbitals $\Psi_{i}(\br t)$, the only physical significance
of which is to build the electronic density according to
\begin{equation}
n(\br t)=\sum_{i=1}^{N}|\Psi_{i}(\br t)|^2.
\label{ks-dens}
\end{equation}
Kohn-Sham Hamiltonian $H_\mathrm{KS}$ and Schr\"odinger Hamiltonian $H$ differ only by an effective potential  in the 
scalar potential $\phi_\mathrm{KS}(\br t)$. This effective potential depends on the Kohn-Sham orbitals themselves, 
$$
\phi(\br t)\,\rightarrow\,
\phi_{\mathrm{KS}}[\Psi_{1}(\br t),\Psi_{2}(\br t),\ldots,
\Psi_{N}(\br t)]\equiv\phi_{\mathrm{KS}}(\br t),
$$
\begin{equation}
H_{\mathrm{KS}}=\left( \frac{1}{2}[\bp+\bA(\br t)]^2+
\phi_{\mathrm{KS}}(\br t)\right).
\label{ks-h}
\end{equation}
Thus, the Kohn-Sham equation corresponds to a set of nonlinear Schr\"odinger equations for the Kohn-Sham orbitals $\Psi_{i}(\br t)$. From the numerical point of view the nonlinearity does not pose a serious problem. The important point is that \reff{s-h} and \reff{ks-h} have the same structure so that the same propagation scheme can be used for both the Schr\"odinger wavefunction and the Kohn-Sham orbitals. In the following we shall present the propagation algorithm employed in the \qprop\ package.

\subsection{\label{ss:propagation-algorithm}Propagation algorithm}

\subsubsection*{Short-time propagator}
Equation (\ref{TDSE}), together with an initial wavefunction $\Psi(\br, t=t_0)$, governs the time evolution of $\Psi(\br t)$ for all times $t$. Introducing the propagator 
\begin{equation}
U(t_2,\ t_1) = \mathrm{T}\exp\left(-\imagi\int_{t_1}^{t_2} H(\tau)\, \diff\tau \right),
\label{arbitrary-time-prop}
\end{equation}
where $\mathrm{T}$ denotes the time-ordering operator, one has
\[ \Psi(\br t_2) = U(t_2,t_1) \Psi(\br t_1) . \]
In most numerical approaches the time-propagation from the initial time $t_0$ to the final one $t_\mathrm{f}$ is divided into small steps $\Delta t$ during which the possibly explicit time-dependence of the Hamiltonian and the time-ordering can be ignored so that the short-time propagator simplifies to 
\begin{equation}
U(t + \Delta t, t) \approx \exp[-\imagi\Delta t H(t+\Delta t/2)].
\label{short-time-prop}
\end{equation}
The finite-time propagation from $t_0$ to $t_\mathrm{f}$ can be decomposed in a product of short-time propagators, 
\begin{equation}
U(t_\mathrm{f},t_0)=\prod_{i=0}^{M-1} U(t_i + \Delta t, t_i)
\end{equation}
with $\Delta t = (t_\mathrm{f}-t_0)/M$.

\subsubsection*{Crank-Nicolson approximation for the short-time propagator}
The short-time propagator \reff{short-time-prop} can be straightforwardly applied to a wavefunction only if the Hamiltonian is diagonal. In order to avoid diagonalization each time step the Crank-Nicolson\footnote{The last name of Phyllis Nicolson is often misspelled ``Nicholson''.} approximant is used \cite{crank:1947}. 
The Crank-Nicolson approximation is easily motivated by the fact that the short-time
propagation of $\Psi(\br t)$ half a timestep forward generates the same 
state as
the propagation of $\Psi(\br, t+\Delta t)$ half a timestep backward,
\begin{equation}
U(t+\Delta t/2, t+\Delta t) \Psi(\br, t+\Delta t)=
U(t+\Delta t/2, t) \Psi(\br t).
\label{step-fwd-bwd}
\end{equation}
Expanding the exponent in the short-time propagator (\ref{short-time-prop})
in a Taylor series, the unitary Crank-Nicolson propagator (accurate to the third order in the time step $\Delta t$)
\begin{equation}
 U(t+\Delta t,  t)=U_\mathrm{CN}(t+\Delta t,  t)+O(\Delta t)^3,\quad U_\mathrm{CN}(t+\Delta t,  t)=\frac{1-\imagi\frac{\Delta t}{2}\, H\left(t+\frac{\Delta t}{2}\right)}
{1+\imagi\frac{\Delta t}{2}\, H\left(t+\frac{\Delta t}{2}\right)}
\label{c-n-appr}
\end{equation}
is obtained. 
The actual implementation of the Crank-Nicolson propagator leads to an implicit algorithm. Because of the Hamiltonian in the denominator we cannot apply $U_\mathrm{CN}$ directly to a wavefunction (unless the Hamiltonian is diagonal). Hence, we go back to (\ref{step-fwd-bwd}) and actually solve
\begin{equation}
(1+\imagi\Delta t H/2) \Psi(\br, t+\Delta t) =
(1-\imagi\Delta t H/2) \Psi(\br t) \label{CN_tobesolved}
\end{equation}
for $\Psi(\br, t+\Delta t)$ each time step. We observe that the Crank-Nicolson propagator
requires the application of the Hamiltonian $H$ to the orbitals 
$\Psi(\br t)$ and $\Psi(\br, t+\Delta t)$. Further below we show how the wavefunction is expanded and discretized, leading to implicit matrix equations corresponding to \reff{CN_tobesolved}.

\subsubsection*{Kohn-Sham Hamiltonian in the linearly polarized laser field}
In this subsection we discuss the more general case of a Kohn-Sham Hamiltonian \reff{ks-h} keeping in mind that the Schr\"odinger Hamiltonian \reff{s-h} is obtained by omitting the electron-electron interaction altogether. 
We write the Kohn-Sham Hamiltonian in the form 
\begin{equation}
H_{\mathrm{KS}}(t)=-\frac{1}{2}\nabla^2+V_I(t) +V(r)+V_\mathrm{ee}[n(\br t)]
\label{ks-ah}
\end{equation}
where,  apart from the familiar kinetic energy term $-\displaystyle\nabla^2/2$,
the Hamiltonian contains the interaction with the electromagnetic field $V_I(t)$ in dipole approximation, the central-field potential of the nucleus (or atomic core) $V(r)$, and the Kohn-Sham effective potential $V_\mathrm{ee}[n(\br t)]$.
In principle, the effective electron-electron interaction potential $V_\mathrm{ee}[n(\br t)]$
can be written as a functional of the total electronic density $n(\br t)$.
In practice, it is approximated using the Kohn-Sham orbitals $\Psi_{i}(\br t)$.
Details on the implemented approximations will be presented below
in Sec.~\ref{ss:one-electron-effective-potentials}.
Here, we only assume that the radial multipole functions $V_\mathrm{ee}^i(rt)$, $i=0,1,2$ in the expansion
\begin{equation}
V_\mathrm{ee}[n(\br t)]=V_\mathrm{ee}^0(rt)+
V_\mathrm{ee}^1(rt)\cos \theta+V_\mathrm{ee}^2(rt)\frac{1}{2}
(3\cos^2 \theta-1)+\cdots 
\label{v_ee-expansion}
\end{equation}
are known. Terms $i>2$, i.e., beyond the quadrupole, are neglected. 
Note that the expansion \reff{v_ee-expansion} holds for linearly polarized light only since the effective potential $V_\mathrm{ee}[n(\br t)]$ is assumed to remain  azimuthally symmetric with respect to the polarization axis.

The interaction with the laser field in dipole approximation is usually established  either in length or velocity gauge. 
Thus the interaction operator $V_I(t)$ for linear polarization $\parallel \be_z$ reads in general 
\begin{equation}
V_I(t)=-\imagi A(t)\frac{\partial}{\partial z}+\frac{A^2(t)}{2}+zE(t), \label{VI}
\end{equation}
where $A(t)\be_z$ is the vector potential and $E(t)\be_z$ is the electric field 
of the laser pulse.  In the velocity gauge only the first two terms are present whereas in the length gauge only the third term is left. The purely time-dependent term $A^2(t)/2$ is easily transformed away and, in fact, is by default not taken into account in \qprop. 
Depending on the  problem, either length or velocity gauge may be advantageous. For instance, high-order above threshold ionization is treated in velocity gauge at lower computational cost than in length gauge \cite{cormier:1996}. The ionization rate of heavy ions, on the other hand, is more easily calculated using length gauge.

\subsubsection*{Radial--angular separation of the Hamiltonian}
Atomic systems in their ground state manifest a prevailing spherical symmetry. 
Expanding the atomic orbitals $\Psi_{i}(\br t)$ in spherical
harmonics $\rY_{lm}(\theta,\varphi)=\rY_{lm}(\Omega)$,
\begin{equation}
\Psi_{i}(\br t)=\frac{1}{r}\sum_{l=0}^\infty\sum_{m=-l}^l \Phi_{ilm}(rt)\rY_{lm}(\Omega),
\label{general-expansion}
\end{equation}
and inserting this expansion into (\ref{TDSE}) with the Kohn-Sham 
Hamiltonian (\ref{ks-ah}) and the dipole interaction for linear polarization \reff{VI}, a set of differential 
equations for the radial orbitals $\Phi_{ilm}(rt)$ is obtained, 
\begin{multline}
\imagi\frac{\partial\Phi_{ilm}(rt)}{\partial t} = 
\left(-\frac{1}{2}\frac{\partial^2}{\partial r^2} + 
V(r) + \frac{l (l + 1)}{2r^2}\right) \Phi_{ilm}(rt) + 
\sum_{l'}\langle\rY_{lm}|V_\mathrm{ee}[n(\br t)]|\rY_{l'm}
\rangle \Phi_{il'm}(rt) +  \\
+ \imagi A(t)\left( -r \sum_{l'} \langle\rY_{lm}|\cos \theta|\rY_{l'm}\rangle
 \frac{\partial }{\partial r} \frac{\Phi_{il'm}(rt)}{r}
+ \sum_{l'} \langle\rY_{lm}|\sin \theta 
\frac{\partial}{\partial \theta}|\rY_{l'm}\rangle 
\frac{\Phi_{il'm}(rt)}{r}  \right)+\\+  
r E(t)\sum_{l'}\langle\rY_{lm}|\cos \theta|\rY_{l'm}\rangle \Phi_{il'm}(rt).
\label{req-1}
\end{multline}
The latter equation shows that the radial orbitals for different magnetic
quantum numbers $m$ are decoupled for a linearly polarized laser field in dipole approximation.
Hence, the general expansion 
(\ref{general-expansion}) can be reduced to 
\begin{equation}
\Psi_{i}(\br t) = \frac{1}{r} \sum_{l=0}^{\infty} 
\Phi_{ilm}(rt) \rY_{lm}(\Omega) \approx 
\frac{1}{r}\sum_{l=0}^{L-1} \Phi_{ilm}(rt) \rY_{lm}(\Omega),
\label{reduced-expansion}
\end{equation}
where the $i$-th orbital possesses a single, well-defined magnetic quantum number $m=m_i$.
Consequently, only $L$ radial orbitals per Kohn-Sham orbital $i$ will be involved in the propagation for linear polarization, whereas
$L^2$ radial orbitals are involved in the general expansion 
(\ref{general-expansion}). The maximum orbital quantum number $l=L-1$ on the numerical grid is related to the number of absorbed photons and must be chosen sufficiently large. The spectral analysis explained in  Sec.~\ref{ss:winop} below can be used to check whether $L$ was chosen properly (see also Sec.~\ref{ss:winop-example}). 

\subsubsection*{Breaking-down the Hamiltonian}
Inserting the electron-electron interaction potential (\ref{v_ee-expansion})
into (\ref{req-1}) and calculating the matrix elements with the
spherical harmonics, \reff{req-1} can be recast in the matrix form \cite{Muller:1999}
\begin{equation}
\imagi\frac{\partial}{\partial t} \bm{\Phi}_i(r t) = 
\left( \bm{\mathrm{H}}_{\mathrm{at}} + 
\bm{\mathrm{H}}_{\mathrm{mix}} + 
\bm{\mathrm{H}}_{\mathrm{ang}}^{(1)} + 
\bm{\mathrm{H}}_{\mathrm{ang}}^{(2)} + 
\bm{\mathrm{H}}_{\mathrm{ang}}^{(3)} \right)\bm{\Phi}_i(r t)
\end{equation}
where $\bm{\Phi}_i(r t)$ denotes a column vector in $l$-space,
\beq
\bm{\Phi}_i(r t) =
[\Phi_{i 0 m}(r t), \Phi_{i 1 m}(r t),\ldots, \Phi_{i (L-1) m}(r t)]^{\mathrm{T}}, \eeq 
and the 
matrices $\bm{\mathrm{H}}_{\mathrm{at}}$, $\bm{\mathrm{H}}_{\mathrm{mix}}$, 
$\bm{\mathrm{H}}_{\mathrm{ang}}^{(1)}$, $\bm{\mathrm{H}}_{\mathrm{ang}}^{(2)}$,
$\bm{\mathrm{H}}_{\mathrm{ang}}^{(3)}$ read
\begin{equation}
\bm{\mathrm{H}}_{\mathrm{at}} = 
-\frac{1}{2}\frac{\partial^2}{\partial r^2} + V(r) + \frac{l(l+1)}{2r^2} + 
V_\mathrm{ee}^0(r t) + p_{l m} V_\mathrm{ee}^2(r t),
\end{equation}

\begin{equation}
\bm{\mathrm{H}}_{\mathrm{mix}} =  -\imagi A(t) 
\begin{pmatrix}
0 & c_{0 m} & 0 & 0 & \ldots \\
c_{0 m} & 0 & c_{1 m} & 0 & \ldots \\
0 & c_{1 m} & 0 & c_{2 m} & \ldots \\
\vdots & 0 & c_{2 m} & 0 & \ldots
\end{pmatrix}   \frac{\partial }{\partial r},
\end{equation}

\begin{equation}
\bm{\mathrm{H}}_{\mathrm{ang}}^{(1)} =  -\imagi \frac{A(t)}{r} 
\begin{pmatrix}
0 & c_{0 m} & 0 & 0 & \ldots \\
-c_{0 m} & 0 & 2 c_{1 m} & 0 & \ldots \\
0 & - 2 c_{1 m} & 0 & 3 c_{2 m} & \ldots \\
\vdots &  0 & - 3 c_{2 m} & 0 & \ldots
\end{pmatrix},
\end{equation}

\begin{equation}
\bm{\mathrm{H}}_{\mathrm{ang}}^{(2)} =  ( r  E(t) + V^1_\mathrm{ee}(r t) ) 
\begin{pmatrix}
0 & c_{0 m} & 0 & 0 & \ldots \\
c_{0 m} & 0 &  c_{1 m} & 0 & \ldots \\
0 &  c_{1 m} & 0 &  c_{2 m} & \ldots \\
\vdots &  0 &  c_{2 m} & 0 & \ldots
\end{pmatrix},
\end{equation}

\begin{equation}
\bm{\mathrm{H}}_{\mathrm{ang}}^{(3)} =  V^2_\mathrm{ee}(r t) 
\begin{pmatrix}
0      & 0 & q_{0m} & 0 & \ldots \\
0      & 0 & 0 & q_{1m} & \ldots \\
q_{0m} & 0 & 0 & 0 & \ldots \\
\vdots & q_{1m} & 0 & 0 & \ldots
\end{pmatrix}
\end{equation}
with
\begin{equation}
c_{lm} = \sqrt{\frac{ ( l + 1 )^2 - m^2 }{
( 2 l + 1 )( 2 l + 3 )}},
\ \ 
p_{lm} = \frac{l (l + 1 ) - 3 m^2}{( 2 l + 1 )( 2 l + 3 )}, \label{coeffcandp}
\end{equation}
\begin{equation}
q_{lm} = \frac{3}{2 ( 2 l + 3 )}
\sqrt{\frac{[ ( l + 1 )^2 - m^2 ][ ( l + 2 )^2 - m^2 ]}
{( 2 l + 1 )( 2 l + 5 )}}. \label{coeffq}
\end{equation}
Note that $\bm{\mathrm{H}}_{\mathrm{at}}$ is diagonal in $l$-space,  $\bm{\mathrm{H}}_{\mathrm{mix}}$, $\bm{\mathrm{H}}_{\mathrm{ang}}^{(1)}$, and $\bm{\mathrm{H}}_{\mathrm{ang}}^{(2)}$ couple an $l$-component $\Phi_{i l m}(r t)$ with $\Phi_{i (l\pm 1) m}(r t)$ while $\bm{\mathrm{H}}_{\mathrm{ang}}^{(3)}$ couples  $\Phi_{i l m}(r t)$ with $\Phi_{i (l\pm 2) m}(r t)$.

The calculation of the spatial derivatives in $\bm{\mathrm{H}}_{\mathrm{at}}$
and $\bm{\mathrm{H}}_{\mathrm{mix}}$ will be discussed below.
Next, the matrix $\bm{\mathrm{H}}_{\mathrm{mix}}$ is split
into a sum over mutually commuting matrices $\bm{\mathrm{H}}_{\mathrm{mix}}^{lm}$, 
\begin{equation}
\bm{\mathrm{H}}_{\mathrm{mix}} \equiv 
\sum_{l=0}^{L-2} \bm{\mathrm{H}}_{\mathrm{mix}}^{lm},
\end{equation}
\begin{multline}
\bm{\mathrm{H}}_{\mathrm{mix}}^{0m} = 
-\imagi A(t) \begin{pmatrix}
0 & c_{0 m} & 0 & 0 & \ldots \\
c_{0 m} & 0 & 0 & 0 & \ldots \\
0 & 0 & 0 & 0 & \ldots \\
\vdots & 0 & 0 & 0 & \ldots
\end{pmatrix}   \frac{\partial }{\partial r}, \\
\bm{\mathrm{H}}_{\mathrm{mix}}^{1m} = 
-\imagi A(t) \begin{pmatrix}
0 & 0 & 0 & 0 & \ldots \\
0 & 0 & c_{1 m} & 0 & \ldots \\
0 & c_{1 m} & 0 & 0 & \ldots \\
\vdots & 0 & 0 & 0 & \ldots
\end{pmatrix}   \frac{\partial }{\partial r} ,  \ldots
\end{multline}
Commutativity of the $\bm{\mathrm{H}}_{\mathrm{mix}}^{lm}$ allows us to use the algebraic equation $\exp(a+b)=\exp(a)\exp(b)$ for the splitting of the corresponding propagator without introducing any additional errors,
\begin{equation}
\exp\left[-\imagi\Delta t   
\left(\sum_{l=0}^{L-2} \bm{\mathrm{H}}_{\mathrm{mix}}^{lm}\right)\right] 
= \prod_{l=0}^{L-2} \exp\left(-\imagi\Delta t   
\bm{\mathrm{H}}_{\mathrm{mix}}^{lm}\right).
\end{equation}
Moreover, only two elements in the matrices 
$\bm{\mathrm{H}}_{\mathrm{mix}}^{lm}$ are different from zero, making them effectively to $2 \times 2$ matrices.
Analogously, the Hamiltonians ${\bm{\mathrm{H}}_{\mathrm{ang}}^{(1)}}$,
${\bm{\mathrm{H}}_{\mathrm{ang}}^{(2)}}$, and 
${\bm{\mathrm{H}}_{\mathrm{ang}}^{(3)}}$ can be decomposed in 
a sum of commuting terms as well.

\subsubsection*{Calculation of the spatial derivatives}
The calculation of the spatial derivatives in $\bm{\mathrm{H}}_{\mathrm{mix}}$ and
$\bm{\mathrm{H}}_{\mathrm{at}}$ is implemented using improved finite 
difference expressions \cite{Muller:1999}. The radial orbitals $\Phi_{ilm}(r t)$
are represented on an equidistant grid with grid spacing $\Delta r=h$ and $N_r$ grid points, 
\begin{equation}
\Phi_{ilm}(r t) = [ \Phi_{ilm}(r_1 t), 
\Phi_{ilm}(r_2 t), \ldots, \Phi_{ilm}(r_{N_r} t) ]^{\mathrm{T}}, \ \ 
r_n = h   n,\ \ \text{} n = 1, 2,\ldots N_r.
\label{grid}
\end{equation}
Implicit fourth order 
Simpson and Numerov expressions for the first and the second
derivative are employed,
\begin{equation}
\pabl{\Phi_{ilm}}{r} = \Phi_{ilm}' \approx
\left( 1 + \frac{h^2}{6} \bm{\Delta}_2 \right)^{-1} 
\bm{\Delta}_1  \Phi_{ilm} =: 
\bm{\mathrm{M}}_1^{-1} \bm{\Delta}_1 \Phi_{ilm},
\label{simpson}
\end{equation}
\begin{equation}
\pabl{^2\Phi_{ilm}}{r^2} = \Phi_{ilm}'' \approx
\left( 1 + \frac{h^2}{12} \bm{\Delta}_2 \right)^{-1} 
\bm{\Delta}_2  \Phi_{ilm} =: -2 
\bm{\mathrm{M}}_2^{-1} \bm{\Delta}_2  \Phi_{ilm}.
\label{numerov}
\end{equation}
The finite difference operators 
$\bm{\Delta}_1 \Phi_{ilm}(r_n t) := 
( \Phi_{ilm}(r_{n+1} t) - \Phi_{ilm}(r_{n-1} t) ) / 2 h$
and $\bm{\Delta}_2 \Phi_{ilm}(r_n t) :=
( \Phi_{ilm}(r_{n+1} t) - 2 \Phi_{ilm}(r_{n} t) + 
\Phi_{ilm}(r_{n-1} t) ) / h^2$ can be written as 
$N_r\times N_r$ matrices, and $\bm{\mathrm{M}}_1$, $\bm{\mathrm{M}}_2$ are given by
\begin{equation}
\bm{\mathrm{M}}_1 = \frac{1}{6} 
\begin{pmatrix}
4 & 1 &   &  \\
1 & 4 & 1 &  \\
   & 1 & 4 & \ddots\\
   &  & \ddots & \ddots
\end{pmatrix}, \ \ \bm{\mathrm{M}}_2 = -\frac{1}{6} 
\begin{pmatrix}
10 & 1 &  &  \\
1 & 10 & 1 &  \\
 & 1 & 10 & \ddots\\
 &  & \ddots & \ddots
\end{pmatrix}.
\end{equation}
In order to ensure unitary time propagation, the matrix $\bm{\mathrm{M}}_1^{-1}\bm{\Delta}_1 $ must be antihermitian. To that end the upper left and lower right corners of $\bm{\Delta}_1$
and $\bm{\mathrm{M}}_1$ are modified \cite{Muller:1999},
\begin{equation}
(\bm{\Delta}_1)_{11} = \frac{y}{2h}, \quad  (\bm{\Delta}_1)_{N_rN_r} = -\frac{y}{2h}, \quad 
(\bm{\mathrm{M}}_1)_{11} = (\bm{\mathrm{M}}_1)_{N_rN_r}  = \frac{4+y}{6}
\end{equation}
with $y=\sqrt{3}-2$. 

In the case of the Coulomb potential $-Z/r$,
the second derivative of a radial $s$-state orbital satisfies at the 
origin $r=0$ the relation 
\beq \Phi_{i(l=0)m}''(0  t) = -2 Z \Phi_{i(l=0)m}'(0  t) \neq 0. \eeq  
This fact can be taken into account by modifying the upper left 
elements in $\bm{\Delta}_2$ and $\bm{\mathrm{M}}_2$ \cite{Muller:1999},
\begin{equation}
(\bm{\Delta}_2)_{11} = -\frac{2}{h^2} 
\left( 1 - \frac{Zh}{12 - 10 Zh} \right),\ \ 
(\bm{\mathrm{M}}_2)_{11} = -2 
\left( 1 + \frac{h^2}{12} (\bm{\Delta}_2)_{11} \right). \label{Zcorrections}
\end{equation}

The fourth order expressions (\ref{simpson}) and (\ref{numerov}), and their modifications to ensure unitary time propagation and the correct behavior at $r=0$, hardly add additional cost to the propagation routine (see \cite{Muller:1999} or the additional documentation provided with \qprop\ and online at \cite{qpropwebpage}).

\subsubsection*{Splitting of the time-propagator}
Summarizing the previous sections, we write the Hamiltonian
\begin{equation}
\bm{\mathrm{H}} = \bm{\mathrm{H}}_{\mathrm{at}} + 
\sum_{l=0}^{L-2} \left( \bm{\mathrm{H}}_{\mathrm{mix}}^{lm} + 
{\bm{\mathrm{H}}_{\mathrm{ang}}^{(1)}}^{lm} + 
{\bm{\mathrm{H}}_{\mathrm{ang}}^{(2)}}^{lm} \right) + 
\sum_{l=0}^{L-3} {\bm{\mathrm{H}}_{\mathrm{ang}}^{(3)}}^{lm},
\label{h-l-r}
\end{equation}
where all terms can be written as a tensor product of operators acting either in ``$l$-space'' or ``$r$-space,'' 
\begin{equation}
\bm{\mathrm{H}}_{\mathrm{at}} = 
\bm{1}_l \otimes 
\left(\bm{\mathrm{M}}_2^{-1}\Delta_2 + V(r) + \frac{l(l+1)}{2r^2}  +  
V_\mathrm{ee}^0(r t) + p_{l m} V_\mathrm{ee}^2(r t)\right),
\end{equation}
\begin{equation}
\bm{\mathrm{H}}_{\mathrm{mix}}^{lm} = 
-\imagi A(t) \bm{\mathrm{L}}^{lm} \otimes 
\bm{\mathrm{M}}_1^{-1} \Delta_1 ,
 \end{equation}
\begin{equation}
{\bm{\mathrm{H}}_{\mathrm{ang}}^{(1,2)}}^{lm} = 
-\imagi A(t) \bm{\mathrm{T}}^{lm} \otimes 
\frac{1}{r} \bm{\mathrm{1}}_r + \bm{\mathrm{L}}^{lm} \otimes 
\left( r E(t) + V_\mathrm{ee}^{1}(r t)\right) \bm{1}_r ,
 \end{equation}
\begin{equation}
{\bm{\mathrm{H}}_{\mathrm{ang}}^{(3)}}^{lm} = \bm{\mathrm{P}}^{lm} \otimes 
V_\mathrm{ee}^{2}(r t) \bm{1}_r.
\end{equation}
$\bm{1}_l$ and $\bm{1}_r$ are unity matrices, and 
$\bm{\mathrm{L}}^{lm}$, $\bm{\mathrm{T}}^{lm}$, and $\bm{\mathrm{P}}^{lm}$
are $2\times2$ matrices
\begin{equation}
\bm{\mathrm{L}}^{lm} = 
\begin{pmatrix}
0 & c_{lm}\\
c_{lm} & 0
\end{pmatrix} ,\quad 
\bm{\mathrm{T}}^{lm} = (l+1) 
\begin{pmatrix}
0 & c_{lm}\\
-c_{lm} & 0
\end{pmatrix} , \quad 
\bm{\mathrm{P}}^{lm} = 
\begin{pmatrix}
0 & q_{lm}\\
q_{lm} & 0
\end{pmatrix}.
\end{equation}
The matrices $\bm{\mathrm{L}}^{lm}$ and $\bm{\mathrm{T}}^{lm}$ act on the $l$th and ($l+1$)th component of the wavefunction while the matrix $\bm{\mathrm{P}}^{lm}$ acts on the $l$th and ($l+2$)th component.

The short-time propagator (\ref{short-time-prop}) with the Hamiltonian
(\ref{h-l-r}) is approximated up to third order in 
$\Delta t=2\tau$ as the product
\begin{multline}
U_{\mathrm{split}}( t + 2\tau, t ) = 
\prod_{l=L-3}^{0}\exp\left(-\imagi \tau {\bm{\mathrm{H}}_{\mathrm{ang}}^{(3)}}^{lm}\right)
\prod_{l=L-2}^{0} \left[ \exp\left(-\imagi \tau {\bm{\mathrm{H}}_{\mathrm{ang}}^{(1,2)}}^{lm}\right)
\exp\left(-\imagi \tau {\bm{\mathrm{H}}^{lm}_{\mathrm{mix}}}\right)\right] \times\\ \times
  \exp(-2 \imagi \tau  \bm{\mathrm{H}}_{\mathrm{at}})\times\\ \times
\prod_{l=0}^{L-2}\left[ \exp\left(-\imagi \tau \bm{\mathrm{H}}^{lm}_{\mathrm{mix}}\right)
\exp\left(-\imagi \tau {\bm{\mathrm{H}}_{\mathrm{ang}}^{(1,2)}}^{lm}\right) \right]
\prod_{l=0}^{L-3}\exp\left(-\imagi \tau {\bm{\mathrm{H}}_{\mathrm{ang}}^{(3)}}^{lm}\right),\label{finalprop}
\end{multline}
where each factor $\exp(-\imagi\tau\cdots)$ is approximated by the corresponding Crank-Nicolson expression (\ref{c-n-appr}). 

Further details about the actual implementation of the propagation algorithm can be found in
the documentation provided with the program and online at \cite{qpropwebpage}.
We only mention here that in \textsc{Qprop} the 
above described propagation procedure for linear polarization using expansion  (\ref{reduced-expansion}),
as well as the more general algorithm for elliptical polarization using expansion (\ref{general-expansion})
are implemented. The propagator in the latter case is even more complex than \reff{finalprop}. It is used in the example in Sec.~\ref{ss:rlm-example}.

\subsection{\label{ss:imag-prop}Imaginary time propagation}

The time-dependent Schr\"o{}dinger equation (\ref{TDSE})
determines the evolution of a quantum system in time,
starting from an initial state $\Psi(\br t_0)$.
Often the initial state
is the ground state of the unperturbed quantum system with
\begin{equation}
H_0 = -\frac{1}{2} \nabla^2  +  V(r).
\label{hamiltonian-wo-field}
\end{equation}
Ground states are analytically known only for a few specific potentials. In contrast, numerical approaches allow to find ground states for arbitrary potentials.  They are often based on the fact
that the total energy of the ground state is minimal.
In \textsc{Qprop} the ground state can be determined using the ordinary propagation algorithm presented above in Sec.~\ref{ss:propagation-algorithm} with, however, the real time step $\Delta t$ replaced by an imaginary time step 
$\Delta t \to -\imagi \Delta t$. Why this works may be seen as follows: 
Let an arbitrary wavefunction $\Psi(\br 0)$ be expanded in eigenstates $\psi_n(\br)$ of the Hamiltonian $H_0$. The wavefunction at later times then reads (without perturbations)
\begin{equation}
\Psi(\br t)  =  \sum_{n} a_n  \exp(-\imagi  \energy_n  t)  \psi_n(\br)
\end{equation}
where $\energy_n$ are the eigenenergies corresponding to $\psi_n(\br)$.
Propagating one imaginary time step leads to
\beq \Psi(\br \Delta t)  =  \sum_{n} a_n  \exp(-  \energy_n  \Delta t)  \psi_n(\br).\eeq
The factor $\exp(-  \energy_n  \Delta t)$ is biggest for minimum $\energy_n=\energy_{\min}$, which means that during imaginary time propagation the corresponding state $\psi_{\min}(\br)$  dies out slowest (if $\energy_{\min}>0$) or explodes fastest (if $\energy_{\min}<0$). The (renormalized) $\Psi(\br t)$ thus converges to the ground state $\psi_{\min}(\br)$.  

The choice of the initial ``guess'' for the ground state wavefunction before 
imaginary time propagation
is not critical but affects the time needed for convergence.
The method even works with a random initial wavefunction,
as it is illustrated in Sec.~\ref{ss:imag-time-prop}. 

Imaginary time propagation can be also applied in the case of several Kohn-Sham orbitals (see example in Sec.~\ref{ss:gs-ne}). The Pauli exclusion principle is implemented via Gram-Schmidt orthogonalization \cite{Chaturvedi-etal:1998} each imaginary time step.

\subsection{\label{ss:imag-pot}Imaginary absorbing potential}
An obvious limitation of propagation in position space is 
the finite size of the numerical grid. Electron density approaching the 
boundaries---if treated without special care---will be reflected and 
may cause undesired, spurious effects. However, the electron density reaching the (sufficiently remote) boundary can be considered to contribute to ionization only but otherwise does not affect the dynamics of the atomic remainder. In order to avoid spurious effects it is desirable and, in fact, possible to formulate exact permeable boundary conditions. Unfortunately, such permeable boundary conditions are easily implemented in one-dimensional calculations only \cite{Boucke:1997}.  In three-dimensional calculations so-called ``absorbing boundaries'' \cite{Neuhauser:1989}, although not mathematically rigorous, proved to be most convenient and practicable. 
In \textsc{Qprop} the 
absorbing boundary is of the form $- \imagi V_{\mathrm{im}}(r)$ with $V_{\mathrm{im}}(r)\geq 0$ to be defined by the user. Clearly, $V_{\mathrm{im}}(r)$ should be a function that is close to zero in the main interaction region and increases up to a sufficiently high value at the grid boundary $R=N_r\Delta r$.  A nonvanishing imaginary potential destroys unitary time propagation and thus leads to ``dissipation'' of the wavefunction. The decreasing norm $N(t)=\vert \langle \Psi(t)\vert \Psi(t)\rangle\vert^2$ may be used to evaluate the ionization probability $P(t)=1-N(t)$ (see example in Sec.~\ref{ss:real-time-prop-h}). 

\subsection{\label{ss:one-electron-effective-potentials}Effective potentials}
According to the Hohenberg-Kohn theorem (see, e.g., \cite{Kohn:1999} and references therein), the electron density determines all observable quantities of a quantum systems, i.e., all observables are functionals of the density. 
Only a few of these functionals are known explicitly while  many of them must be approximated in practice. 

An electron density $n(\br)$ in an external potential $V(\br)$ yields  the potential energy
\begin{equation}
E_{\mathrm{pot}}[n] = \int V(\br) n(\br)\, \diff^3 r,
\label{ext-energy}
\end{equation}
while the exact kinetic energy as an explicit functional of the density $T_{\mathrm{kin}}[n(\br)]$ is, in general,  unknown.

Kohn and Sham (see, e.g., \cite{Kohn:1999} and references therein) considered an auxiliary system of noninteracting
electrons whose density $n_\mathrm{s}(\br)$ coincides with the density $n(\br)$ of the real, interacting system. 
Because the auxiliary electrons do not interact with each other (but move in a common, effective potential),
their Schr\"o{}dinger equation separates, and the kinetic energy
is simply the sum of one-particle kinetic energies,
\begin{equation}
T_\mathrm{s} = - \frac{1}{2} 
\sum_{i}\langle\Psi_{i} | \nabla^2 | \Psi_i\rangle.
\label{kin-energy}
\end{equation}
The Kohn-Sham orbitals $\Psi_{i}(\br)$ satisfy the
Kohn-Sham equation
\begin{equation}
\left[-\halb\nabla^2 + V_\mathrm{ee}(\br) \right]\Psi_{i}(\br) = 
\epsilon_i \Psi_i(\br).
\label{kse}
\end{equation}
The only physical significance of the Kohn-Sham orbitals is to form 
the correct density
\begin{equation}
n_\mathrm{s}(\br) = \sum_{i}|\Psi_i(\br)|^2,
\end{equation}
which, in turn, uniquely determines the potential $V_\mathrm{ee}(\br)$, commonly split in the form
\begin{equation}
V_\mathrm{ee}(\br) = V(\br) + U(\br) + V_{\mathrm{xc}}(\br),
\end{equation}
where $V(\br)$ denotes the same potential that is present in the interacting system's Hamiltonian (e.g., $V(r)=-Z/r$), 
\beq U(\br)=\int \frac{n(\br')}{\vert\br-\br'\vert}\,\diff^3 r' \label{hartreepotential}\eeq 
is the Hartree-potential 
accounting for the mutual repulsion of the electrons, and $V_{\mathrm{xc}}(\br)$
denotes the so-called exchange-correlation potential. The nuclear potential $V(\br)$, although entirely independent of the electron density, can be formally written as the variational derivative of the 
external potential energy (\ref{ext-energy}),
\begin{equation}
V(\br) =  \frac{\delta E_{\mathrm{pot}}[n]}{\delta n(\br)}. \label{ext-pot}
\end{equation}
Analogously, the Hartree potential is the variational derivative of the 
electron-electron repulsion energy,
\begin{equation}
U(\br) = \frac{\delta E_{\mathrm{H}}[n]}{\delta n(\br)}, \qquad    E_{\mathrm{H}}[n]=\frac{1}{2}\int\!\!\int \frac{n(\br) n(\br')}{| \br - \br' |} \,\diff^3r'\,\diff^3 r.
\label{H-pot}
\end{equation}
The exchange-correlation potential $V_{\mathrm{xc}}(\br)$ is split
further into exchange and correlation potential, $V_{\mathrm{xc}}(\br) = 
V_{\mathrm{x}}(\br) + V_{\mathrm{c}}(\br)$.
In what follows, the correlation potential $V_{\mathrm{c}}(\br)$ is neglected.

\subsubsection*{Approximations for time-dependent effective potentials}
The time-independent Kohn-Sham equation (\ref{kse}) has its 
time-dependent counterpart, as already anticipated in Eqs.\ (\ref{TDSE}), (\ref{ks-h}), and \reff{ks-ah}.
Analogous to the Hohenberg-Kohn theorem, the Runge-Gross theorem \cite{runge:1984} asserts
that also in the time-dependent case, the density $n(\br t)$ 
determines all observable quantities. 

In \textsc{Qprop} the time-dependent potentials 
of electron-electron repulsion and exchange are calculated
adopting an adiabatic viewpoint,
i.e., expressions known from stationary density functional theory are evaluated using the now time-dependent density $n(\br t)$.  
The effective potential \reff{v_ee-expansion} is written as a sum of Hartree and exchange potential,
\begin{equation}
V_\mathrm{ee}[n(\br t)] = U[n(\br t)] + V_{\mathrm{x}}[n(\br t)].
\label{u+x}
\end{equation}
The Hartree potential is expanded analogously to (\ref{v_ee-expansion}),
\begin{equation}
U[n(\br t)] = U^0(r t) + 
U^1(r t) \cos \theta + U^2(r t) \frac{1}{2} 
( 3\cos^2 \theta - 1 ) +\cdots .
\label{u-expansion}
\end{equation}
In the current version of \textsc{Qprop} the latter expansion is terminated (at latest) 
after the quadrupole term.  The exchange potential $V_{\mathrm{x}}[n(\br t)]$
is approximated using the expression proposed by Krieger, Li, and Iafrate (KLI)  in \cite{Krieger-etal:1992}. Within the KLI approximation, going beyond the groundstate, monopole term turns out to be hard while in local density approximation it is simple  to go up to the quadrupole term as well.

\subsubsection*{Hartree potential}
The Hartree-potential (\ref{hartreepotential})
is  calculated using the Kohn-Sham density (\ref{ks-dens}) and the orbitals (\ref{reduced-expansion}). Expanding 
${|\br-\br'|}^{-1}$ in spherical harmonics,
the following expressions for the Hartree multipoles
$U^k(r t)$ in equation (\ref{u-expansion}) are obtained,
\begin{equation}
U^0(r t) = \int \frac{1}{r_{>}} \Lambda(r' t)\, \diff r',
\label{hartree0}
\end{equation}
\begin{equation}
U^1(r t) = \int  \frac{r_{<}}{r_{>}^2} \Theta(r' t)\, \diff r',
\label{hartree1}
\end{equation}
\begin{equation}
U^2(r t) = \int  \frac{r_{<}^2}{r_{>}^3} \Xi(r' t)\, \diff r'
\label{hartree2}
\end{equation}
where $r_<$ and $r_>$ are $\min(r,r')$ and $\max(r,r')$, respectively.
The auxiliary functions $\Lambda(r t)$, $\Theta(r t)$, and $\Xi(r t)$ read 
\begin{equation}
\Lambda(r t) = 2 \sum_{i l} | \Phi_{ilm}(r t) |^2,
\label{Lambda}
\end{equation}
\begin{equation}
\Theta(r t) = 2 \sum_{i l} \left[ 
c_{l-1,m} \Phi_{i,l-1,m}^{*}(r t) + 
c_{lm} \Phi_{i,l+1,m}^{*}(r t) \right] \Phi_{ilm}(r t),
\label{Theta}
\end{equation}
\begin{equation}
\Xi(r t) = 2 \sum_{i l} \left[ 
p_{lm} \Phi_{ilm}^{*}(r t) + 
q_{lm} \Phi_{i,l+2,m}^{*}(r t) + 
q_{l-2,m} \Phi_{i,l-2,m}^{*}(r t) \right] \Phi_{ilm}(r t).
\label{Xi}
\end{equation}
The factor $2$ in these expressions stems from the spin degeneracy, i.e., we assume spin-unpolarized systems where each Kohn-Sham orbital is occupied by two electrons of opposite spin. The coefficients $c_{lm}$, $p_{lm}$, and $q_{lm}$ are given by \reff{coeffcandp} and \reff{coeffq}.

\subsubsection*{Krieger-Li-Iafrate approximation to the exchange potential}
It is possible to derive an integral equation that implicitly defines the so-called optimized effective potential \cite{Sharp:1953,Talman:1976,Grabo:2000}. However, this integral equation is difficult to solve, and simpler expressions of comparable accuracy are needed in practice. Krieger, Li, and Iafrate (KLI) \cite{Krieger-etal:1992} simplified the full integral equation and obtained for the exchange potential 
\begin{equation}
V_{\mathrm{x}\sigma}^{\mathrm{KLI}}(\br) = 
V_{\mathrm{x}\sigma}^{\mathrm{Slater}}(\br)  + 
\sum_{i=1}^{N_\sigma-1} \frac{|\Psi_{i\sigma}(\br)|^2}{n_\sigma(\br)} \int
 |\Psi_{i\sigma}(\br')|^2 \left( V_{\mathrm{x}\sigma}^{\mathrm{KLI}}(\br') - 
u_{\mathrm{x} i\sigma}(\br') \right) \, \diff^3 r'
\label{KLI-eq}
\end{equation}
where $\sigma=\uparrow,\downarrow$ denotes the spin variable of  $N_\sigma$ electrons,
\begin{equation}
n_\sigma(\br) = \sum_{i=1}^{N_\sigma} | \Psi_{i\sigma}(\br) |^2 
\end{equation}
is the spin density,
\begin{equation}
V_{\mathrm{x}\sigma}^{\mathrm{Slater}}(\br) = 
\sum_{i=1}^{N_\sigma} \frac{|\Psi_{i\sigma}(\br)|^2}{n_\sigma(\br)} u_{\mathrm{x} i\sigma}(\br) 
\end{equation}
is the so-called Slater-potential, and
\begin{equation}
u_{\mathrm{x} i\sigma}(\br) = \frac{1}{\Psi_{i\sigma}^*(\br)} \frac{\delta E_\mathrm{x}[\{\Psi_{j\sigma}\}]}{\delta \Psi_{i\sigma}(\br)} = -\sum_{j=1}^{N_\sigma} \frac{\Psi_{j\sigma}(\br)}{\Psi_{i\sigma}(\br)} 
\int \frac{\Psi_{i\sigma}^*(\br') 
\Psi_{j\sigma}(\br')}{| \br - \br' |}\, \diff^3 r'
\end{equation}
with the exact exchange energy
\begin{equation}
E_\mathrm{x}[\{\Psi_{j\sigma}\}] = -\halb \sum_{\sigma=\uparrow,\downarrow} \sum_{i,j=1}^{N_\sigma} \int\!\!\int \frac{\Psi_{i\sigma}^*(\br) \Psi_{j\sigma}^*(\br')\Psi_{j\sigma}(\br)\Psi_{i\sigma}(\br')}{|\br-\br'|}\, \diff^3 r' \, \diff^3 r . \label{x-energ}
\end{equation}

The integral equation (\ref{KLI-eq}) can be solved for $V_{\mathrm{x}\sigma}^{\mathrm{KLI}}(\br)$ by  multiplying 
both sides with $|\Psi_{j\sigma}(\br)|^2$ and integrating over space. Introducing the short-hand notation $\langle A \rangle_{j\sigma}$ for the orbital average $\int | \Psi_{j\sigma} (\br)|^2 A(\br)\, \diff^3r$ of an entity $A$, the matrix equation
\begin{equation}
\sum_{i=1}^{N_\sigma-1} ( \delta_{ji} - \mathrm{M}_{ji\sigma} ) Q_{i\sigma} = 
\langle V_{\mathrm{x}\sigma}^{\mathrm{Slater}} -  u_{\mathrm{x} j\sigma}\rangle_{j\sigma}
\label{Q-eq}
\end{equation}
for the $N_\sigma - 1$ coefficients 
\beq
Q_{i\sigma} = \langle V_{\mathrm{x}\sigma}^{\mathrm{KLI}} - 
u_{\mathrm{x} i\sigma} \rangle_{i\sigma} \eeq 
is obtained. The $(N_\sigma - 1)\times(N_\sigma - 1)$ matrix $\mathrm{M}_{ji\sigma}$
is given by
\begin{equation}
\mathrm{M}_{ji\sigma} = \int  
\frac{|\Psi_{j\sigma}(\br)|^2 |\Psi_{i\sigma}(\br)|^2}{n_\sigma(\br)}.
\end{equation}
The term with the highest occupied spin orbital $i=N_\sigma$ is excluded from the sums in \reff{KLI-eq} and \reff{Q-eq} since it can be shown that $Q_{N_\sigma\sigma} = 0$ \cite{Krieger-etal:1992}. 
After solving \reff{Q-eq} for $Q_{i\sigma}$, all the quantities on the right hand
side of (\ref{KLI-eq}) are determined, and the KLI potential can be evaluated.

Currently, only spin-unpolarized systems where $n(\br)= 2 n_\sigma(\br)$, $N_\sigma=N/2$ with $N$ the number of electrons in the system, and $V_{\mathrm{x}\sigma}^{\mathrm{KLI}} = V_{\mathrm{x}}^{\mathrm{KLI}}$ are tractable with \qprop. Orbital degeneracies $>2$ are possible, so that no unnecessary overhead arises for Kohn-Sham orbitals that evolve identically in a laser field. For instance, the four $2p$ orbitals with $\sigma=\uparrow,\downarrow$ and $|m|=1$ behave identically in dipole approximation and thus can be subsumed under a single orbital $\Psi_i(\br t)$ of degeneracy $d_i=4$.

Further computational details can be found in the documentation provided with the program and online \cite{qpropwebpage}.
In the current version of \qprop\ the KLI potential
is restricted to the ground state monopole contribution ${V_{\mathrm{x}}^{\mathrm{KLI}}}^0(r)$. The latter is sufficient to determine state-of-the-art effective potentials for the ground state of spherically symmetric systems (or other systems in central field approximation). The ground state KLI potential  ${V_{\mathrm{x}}^{\mathrm{KLI}}}^0(r,t=0)$ may then be used for either frozen-core calculations or it may be re-calculated each time step as an approximation to the real TDDFT exchange potential. The latter approximation should be acceptable  as long as most of the electrons stay inside the atom and the deviation  from spherical symmetry remains small. The dipole contribution to the time-dependent KLI potential would be already very complicated due to the coupling of many angular momenta. Within simpler approaches (e.g., local density approximation) where the exchange potential is given explicitly as a functional of the density $n(\br t)$, an expansion up to the quadrupole is much simpler.

\subsection{\label{ss:winop}Spectral analysis: the window-operator technique}
\textsc{Qprop} allows to calculate an initial state of
interest and propagates this state in time. By the end of
propagation---usually at the end of the laser pulse---the final state $\Psi(\br)=\Psi(\br,t=t_\mathrm{f})$ is obtained.
It is useful to analyze its spectrum with respect to the Hamiltonian $H_0$ of the atomic system (without laser) whose eigenstates we denote by $\psi_\energy(\br)$,
\begin{equation}
| \Psi \rangle = \sum\!\!\!\!\!\!\!\!\int\ c_{\energy} | \psi_{\energy}\rangle\, \diff\energy.
\end{equation}
In order to avoid the explicit calculation of all the eigenstates $| \psi_{\energy}\rangle$, a window-operator technique very similar to the one  proposed in
\cite{Schafer-and-Kulander:1990,Schafer:1991} is employed. 
A window-operator $W_{\mathcal{E} \gamma n}$ of energy $\mathcal{E}$ and 
of energy width $\gamma$ is defined as
\begin{equation}
W_{\mathcal{E} \gamma n} = 
\frac{\gamma^{2^n}}{(H_0 - \mathcal{E})^{2^n}  +  \gamma^{2^n}}, \label{window}
\end{equation}
where $n$ is the integer order of the window-operator. The higher the order $n$, the more rectangular is the window profile. 

When acting on a state $|\Psi\rangle$, the window-operator 
returns the energy component $|\chi_{\mathcal{E}\gamma n}\rangle$ the scalar product of which gives a measure for the population of states that lie within the window of width $\gamma$, centered around the electron energy $\mathcal{E}$,
\begin{equation}
\langle \chi_{\mathcal{E}\gamma n}|\chi_{\mathcal{E}\gamma n}\rangle = 
\langle \Psi | W_{\mathcal{E}\gamma n}^2 | \Psi\rangle = 
\sum\!\!\!\!\!\!\!\!\int\   \vert c_{\energy'}\vert^2  \left(
\frac{\gamma^{2^n}}{(\energy' -  \mathcal{E})^{2^n} + \gamma^{2^n}}
\right)^2 \, \diff \energy'.
\end{equation}

Hence, for $\gamma \rightarrow 0$ the modulus squared of the energy component
$|\chi_{\mathcal{E}\gamma n}\rangle$ equals $|c_\mathcal{E}|^2$. Finite energy width $\gamma$, on the other hand,  allows to model realistic measurements with a finite energy resolution.

Since $W_{\mathcal{E}\gamma n}$ has the operator $H_0$ in the denominator, $| \chi_{\mathcal{E}\gamma n}\rangle$ is actually calculated by solving
the equation $W_{\mathcal{E}\gamma n}^{-1}  | \chi_{\mathcal{E}\gamma n}\rangle 
= | \Psi\rangle$ using the factorization
\begin{equation}
(H_0 - \mathcal{E})^{2^n}  +  \gamma^{2^n} = 
\prod_{k=1}^{2^{n-1}} \left[H_0 - \mathcal{E} + 
\gamma\exp(\imagi \nu_{nk})\right]
\left[H_0 - \mathcal{E} - \gamma\exp(\imagi \nu_{nk})\right],
\label{fct}
\end{equation}
where the phases $\nu_{n k}$ are uniformly distributed between $0$ and $\pi/2$,
\begin{equation}
\nu_{n k} = ( 2 k - 1 ) \pi / 2^n.
\end{equation}

In the current version of \qprop, the energy component $|\chi_{\mathcal{E}\gamma n}\rangle$ is calculated 
for the fixed order $n = 3$ for both the general
(\ref{general-expansion}) and the reduced expansion (\ref{reduced-expansion}).
$| \chi_{\mathcal{E}\gamma n}\rangle $ is then obtained as an expansion in
spherical harmonics with the radial energy components denoted by $R_{lm}^{\chi}(r)$. The energy components $| \chi_{\mathcal{E}\gamma n}\rangle$ in form of
expansions over spherical harmonics allow to
calculate differential probabilities of electron emission.
The probability 
$P_{\gamma n}(\mathcal{E})$, differential in energy, reads
\begin{multline}
P_{\gamma n}(\mathcal{E}) = 
\langle\chi_{\mathcal{E}\gamma n}| \chi_{\mathcal{E}\gamma n}\rangle = 
\int \diff r\, \diff\Omega\, \left| \sum_{lm} R_{lm}^{\chi}(r) \rY_{lm}(\Omega)\right|^2 = 
\sum_{lm}\int \diff r\, R_{lm}^{\chi *}(r) R_{lm}^{\chi}(r).
\label{pgne}
\end{multline}
Useful information about the emission probability differential in energy and in angle is obtained by simply omitting the integration over the solid angle,
\begin{equation}
P_{\gamma n}(\mathcal{E},  \Omega) = 
\int \diff r\,  \left| \sum_{lm} R_{lm}^{\chi}(r) \rY_{lm}(\Omega) \right|^2.
\label{pgneo}
\end{equation}

The probabilities $P_{\gamma n}(\mathcal{E})$ and 
$P_{\gamma n}(\mathcal{E},\, \Omega)$ simplify for the reduced
expansion (\ref{reduced-expansion}) since the sum over $m$ is absent then. 
Examples for the calculation of electron spectra are given in Secs.~\ref{ss:winop-example} and \ref{ss:rlm-example}.

\section{\label{s:program-structure}Program structure}
The \textsc{Qprop} package is arranged as a library of classes whose objects
represent orbitals, grids, and Hamiltonians. 
In order to use the
library, an executable program has to be written. This program is usually 
short and may be regarded as an extended input-file which
profits from all the powerful \textsc{C++} features. Compared to a foolproofed approach where the user never makes direct contact with the actual data structures, our library-oriented approach requires more knowledge of the internal structures. On the other hand, once the user got acquainted with the important classes and functions,  he or she may really benefit from the huge flexibility. 

Content and functionality of the internal data structures will be 
considered in the following Sec.~\ref{s:internal-structures}. 
The five examples in Sec.~\ref{s:test-calculations} are aimed
at facilitating the first steps in the usage of \qprop.

\subsection{\label{s:internal-structures}Internal data structures}
The classes \texttt{fluid}, \texttt{wavefunction},
\texttt{grid}, and \texttt{hamop} build 
the core of the \textsc{Qprop} library. Objects of classes \texttt{fluid}
and \texttt{wavefunction} represent real valued and
complex valued one-dimensional arrays, respectively.\footnote{The name ``{\tt fluid}'' may appear strange. It is mainly for ``historical'' reasons since this class was originally developed for a fluid code that naturally deals with real valued arrays.}  
The methods provided by these classes allow to initialize the corresponding arrays, to manipulate them, and to store (load) them to (from) files. From the physical point of view, objects of class \texttt{wavefunction} typically represent radial wavefunctions or sets of radial orbitals to be propagated. However, objects of class \texttt{wavefunction} may be used for any auxiliary quantity that can be represented by a complex vector. The heart of the \textsc{Qprop} library, namely the actual short-time propagation \reff{finalprop}, is implemented as a member function of class \texttt{wavefunction}.

The entries of an object of class \texttt{wavefunction} are accessible through an object of class \texttt{grid}, which defines the number of spatial grid points, the number of angular momentum quantum numbers, the number of orbitals, the grid spacing, and whether expansion \reff{general-expansion} or \reff{reduced-expansion} is used. In other words, objects of class \texttt{grid} define the numerical grid on which the objects of class \texttt{wavefunction} are defined, e.g., the number of grid points $N_r$ in radial direction, the upper limit $L-1$ for the $l$ quantum numbers in the expansions
(\ref{general-expansion}) or (\ref{reduced-expansion}), and the number of Kohn-Sham
orbitals.
The most important functions of class
\texttt{grid} are collected in Tab.~\ref{t:grid}.

\begin{table}[htb]
\begin{small}
\caption{\label{t:grid}
Main functions (methods) of class \texttt{grid}.}
\begin{center}
\begin{tabular}{ll p{7cm}} \\[-0.6cm]    
\hline \hline  \\[-0.4cm]
Method                           & Arguments & Description and comments 
\\[0.1cm]  \hline  \\[-0.2cm]
\texttt{set\_dim}
& \texttt{(long dim)} & 
Defines type of expansion, i.e., either
(\ref{general-expansion}) {\tt dim=44} or (\ref{reduced-expansion}) {\tt dim=34}.
\\
\texttt{index}
& \texttt{(long r, long l, long m, long n)} & 
Calculates the index of a wavefunction (or orbital) entry; radial position {\tt r}, angular momentum {\tt l}, magnetic quantum number {\tt m}, orbital no.\ {\tt n}; {\tt n} is irrelevant for {\tt dim=44} while  {\tt m} is irrelevant for {\tt dim=34}
\\
\texttt{set\_delt}
& \texttt{(double dr)} & 
Defines $\Delta r$
\\
\texttt{set\_ngps}
& \texttt{(long N\_r, long L, long N\_orb)} & 
Defines $N_r$, $L$, and 
number of Kohn-Sham orbitals.
\\
\texttt{size}
& \texttt{(void)} & 
Returns size of grid object
\\
\texttt{r}
& \texttt{(long rindex)} 
& Calculates $r$, given {\tt rindex} $\in [0,N_r-1]$
\\ 
\texttt{ngps\_x}
& \texttt{(void)} 
& Returns number of radial gridpoints $N_r$ 
\\
\texttt{ngps\_y}
& \texttt{(void)} 
& Returns number of angular momenta $L$ 
\\
\texttt{ngps\_z}
& \texttt{(void)} 
& Returns number of orbitals $N$ 
\\
\hline \hline 
\end{tabular}\end{center}
\end{small}
\end{table}

Objects of class \texttt{hamop} collect a number of external potentials that set up the Hamiltonian. These potentials are to be defined by the user and are listed in Tab.~\ref{t:hamop}. 
\begin{table}[htb]
\begin{small}
\caption{\label{t:hamop}
Functions needed to define an object of class {\tt hamop}. The argument {\tt me} is for parallel computing purposes where it allows to let the potentials depend on the job number.} 
\begin{center}
\begin{tabular}{ll p{7cm}} \\[-0.6cm]    
\hline \hline  \\[-0.4cm]
Function                           & Arguments & Description and comments 
\\[0.1cm]  \hline  \\[-0.2cm]
{\tt vecpot\_x} & {\tt (double t, int me)} & $x$-component of the vector potential (only relevant for {\tt dim=44}, i.e., with Eq.~(\ref{general-expansion}))
\\
{\tt vecpot\_y} & {\tt (double t, int me)} & $y$-component of the vector potential (only relevant for {\tt dim=44}, i.e., with Eq.~(\ref{general-expansion}))
\\
{\tt vecpot\_z} & {\tt (double t, int me)} & $z$-component of the vector potential (only relevant for {\tt dim=34}, i.e., with Eq.~(\ref{reduced-expansion}))
\\
{\tt scalarpotx} & {\tt (double x, y, z, t, int me)} & Spherically symmetric potential $V(r)$ in \reff{req-1}, {\tt x} corresponds to $r$, {\tt y} and  {\tt z} are not needed
\\
{\tt scalarpoty} &  {\tt (double x, y, z, t, int me)} &  $\equiv 0$ since not used here
\\
{\tt scalarpotz} &  {\tt (double x, y, z, t, int me)} &  $\equiv 0$ since not used here
\\
{\tt field} & {\tt (double t, int me)} & electric field $E(t)$ in  \reff{req-1} \\
{\tt imagpot} & {\tt (long xindex, yindex, zindex,}\\
& \ {\tt double t, grid g)} & imaginary potential $V_\mathrm{im}(r)$ from Sec.~\ref{ss:imag-pot}; {\tt xindex} is the radial index for {\tt dim=34} or {\tt 44} while {\tt yindex} and {\tt zindex} are not relevant here\\
\hline \hline 
\end{tabular}\end{center}
\end{small}
\end{table}

Class \texttt{wavefunction} is clearly the most important part of the \qprop\ library.
As already mentioned, it contains the methods to 
initialize, to load and to store the radial 
orbitals, to propagate them in time, to calculate the observable quantities 
and 
effective potentials. Table \ref{t:functions} shows some of the public methods
to perform these tasks. Several methods are overloaded, i.e.,  they are distinguished by the different sets of arguments only. The meaning of most arguments is self-explanatory.
Since many of the methods need to know in which order the radial
orbitals are organized in the internal, one-dimensional array,\footnote{There are only one-dimensional arrays internally.} the first argument often is of class \texttt{grid}. In order to control the verbosity of 
some complex methods,
there is an integer \texttt{iv} argument at the end of the parameter list.
Setting it to zero suppresses any output to the \texttt{stdout} stream.

\begin{sidewaystable}
\begin{small}
\caption{\label{t:functions}
Selected functions (methods) of class \texttt{wavefunction}.}
\begin{center}
\begin{tabular}{p{4.2cm} p{7.0cm} p{12.3cm}} \\[-0.6cm]    
\hline \hline  \\[-0.4cm]
Method                           & Arguments & Description and comments 
\\[0.1cm]  \hline  \\[-0.2cm]
\texttt{calculate\_staticpot}
& \texttt{(grid g, hamop hamil)} & 
Calculates static part of total potential ({\tt scalarpotx} + centrifugal potential + {\tt imagpot}) using Hamiltonian \texttt{hamil}
                                    \\
\texttt{calculate\_Lambda}
& \texttt{(grid g, fluid deg)} & 
Calculates auxiliary quantity $\Lambda(r t)$ (\ref{Lambda})
                                    \\
\texttt{calculate\_Theta}
& \texttt{(grid g, fluid deg, fluid ms)} & 
Calculates auxiliary quantity $\Theta(r t)$ (\ref{Theta})
                                    \\
\texttt{calculate\_Xi}
& \texttt{(grid g, fluid deg, fluid ms)} & 
Calculates auxiliary quantity $\Xi(r t)$ (\ref{Xi})
                                    \\
\texttt{calculate\_hartree\_zero}
& \texttt{(grid g, fluid Lambda)} & 
Calculates monopole term (\ref{hartree0}) in expansion (\ref{u-expansion})
                                    \\
\texttt{calculate\_hartree\_one}
& \texttt{(grid g, wavefunction Theta)} & Calculates dipole term
(\ref{hartree1}) in expansion (\ref{u-expansion})
                                    \\
\texttt{calculate\_hartree\_two}
& \texttt{(grid g, wavefunction Xi)} & 
Calculates quadrupole term (\ref{hartree2}) in expansion (\ref{u-expansion})
                                    \\
\texttt{calculate\_kli\_zero}
& \texttt{(grid g, fluid Lambda, fluid ls, fluid ms, fluid deg, 
\newline int slateronly, int iv)}
& 
Calculates groundstate KLI potential (\ref{KLI-eq}). Auxiliary
array \texttt{Lambda} must be provided (see function 
\texttt{calculate\_Lambda()} in this table). Flag \texttt{slateronly=1}
leaves only first summand $V_{\mathrm{x}\sigma}^{\mathrm{Slater}}(\br)$
on right hand side of (\ref{KLI-eq})
                                    \\
\texttt{dump\_to\_file\_sh}
& \texttt{(grid g, FILE file, int st, int iv)} & 
Saves orbitals in \texttt{file} 
                                    \\
\texttt{energy}
& \texttt{(double time, grid g, hamop hamil, int me,
  wavefunction staticpot, double charge)} & 
Calculates kinetic energy plus potential energy due to static potential
\texttt{staticpot}, e.g., bracket in equation (\ref{sp}) below
                                    \\
\texttt{expect\_z}
& \texttt{(grid g)} & Calculates $\langle z\rangle=\langle\Psi(t) | z |\Psi(t)\rangle$ for a single orbital.\\
\texttt{expect\_z}
& \texttt{(grid g, fluid deg, const fluid ms)} & Calculates $\langle z\rangle$ for the entire set of Kohn-Sham orbitals.\\
\texttt{init}      & \texttt{(long size)}   & Various ways to initialize {\tt wavefunction} objects (cf.\ examples in Sec.~\ref{s:test-calculations}) 
                                    \\
& \texttt{(grid g, int inittype, double w, fluid ls)} 
                                    \\
& \texttt{(grid g, FILE file, int ooi, int iv)} \\
{\tt init\_rlm} & {\tt (grid g, int inittype, double w, fluid ls, fluid ms)} \\

\texttt{propagate}      & \texttt{(complex timestep, 
double time, grid g, hamop hamil, int me,
wavefunction staticpot, int m, double nuclear\_charge)} & 
Propagates a wavefunction from \texttt{time}
to \texttt{time+timestep}. \texttt{hamil} determines only external
fields in this function. Potential of nucleus is supposed to be contained  
in \texttt{staticpot} (see \texttt{calculate\_staticpot()} in this table).
\texttt{nuclear\_charge} is used to account for correction \reff{Zcorrections}. There is an overloaded version for the propagation of several Kohn-Sham orbitals (cf.\ Sec.~\ref{ss:gs-ne})
                                     \\
\texttt{totalenergy\_hartree}
& \texttt{(grid g, fluid Lambda, fluid U\_0)} & 
Calculates Hartree energy $E_\mathrm{H}[n]$ (\ref{H-pot})
                                    \\
\texttt{totalenergy\_exact\_x}
& \texttt{(grid g, fluid ells, fluid deg)} & 
Calculates exchange energy $E_\mathrm{x}$ \reff{x-energ}
                                    \\
\\[0.1cm]
\hline \hline 
\end{tabular}
\end{center}
\end{small}
\end{sidewaystable}

Objects of class \texttt{fluid} are basically real valued, one-dimensional arrays, i.e., the real counter parts of {\tt wavefunction} objects. They are used to store effective potentials and auxiliary
quantities that are not complex. For instance, {\tt calculate\_hartree\_zero} in Tab.~\ref{t:functions} returns an object of class {\tt fluid}. Usage of the class is 
straightforward and can be understood
from the examples in Sec.~\ref{s:test-calculations}.

Apart from the classes \texttt{wavefunction}, \texttt{grid}, 
\texttt{hamop} and \texttt{fluid}, another class \texttt{cmatrix}
and several functions are defined. Class \texttt{cmatrix} deals with 
matrix operations, some of which  make use of the \texttt{blas} and \texttt{lapack} libraries.
However, class \texttt{cmatrix}  is currently used only inside member functions of class \texttt{wavefunction} so that there is no need to discuss its features in the present paper.

\subsection{\label{s:used-libraries}External libraries and source codes}
\textsc{Qprop} needs the libraries \textsc{blas}, \textsc{lapack}, and  \textsc{f2c}--- all available free of charge \cite{blas, lapack, f2c}. Note that they are part of many {\sc Linux} distributions. 
Apart from the libraries, \qprop\ profits from two programs to be cited:
Clebsch-Gordan coefficients (needed for the KLI potential \reff{KLI-eq}) are calculated using the program \textsc{ned} by Arturo Sierra \cite{Sierra:ned}, and spherical harmonics (needed for the implementation of \reff{pgneo}) are calculated as in the {\sc Racah} package \cite{fritz:2003}.

\subsection{Distribution and installation}
The program is distributed as a single tarball \texttt{qprop.tar.gz}.
The archive is organized in several
subdirectories with source code, makefiles and  \texttt{README}-files.
Installation consists of extracting the tarball and creating links 
to the libraries \textsc{blas}, \textsc{lapack}, and \textsc{f2c} in the subdirectory \texttt{lib/i386/}.
The top directory contains a makefile \texttt{qprop/GNUmakefile.tmpl},
which controls the compilation of the \textsc{qprop} library.
The user will probably need to adjust the variable \texttt{ROOT} in this file.
This variable points to the absolute location of the \textsc{Qprop} package.

\begin{figure}
\begin{verbatim}
 qprop/
       doc/
       lib/i386/
       obj/i386/
       src/
                base/
                circ/
                hydrogen/
                ionization/
                neon/
                winop/
\end{verbatim}
\caption{\label{f:ds}
Directory structure of the \textsc{Qprop} package.}
\end{figure}

Figure \ref{f:ds} shows the directory structure of the \textsc{Qprop}
package. The basic source code is placed in the subdirectory
\texttt{src/base/}. The code of several examples is placed in
subdirectories in \texttt{src/}. Five examples are provided and discussed
in Sec.~\ref{s:test-calculations}.

The other directories contain extra documentation (in \texttt{doc/}),
object files \texttt{.o} (in \texttt{obj/i386/}),  and the library files (or links to)
\texttt{libqprop.a}, \texttt{libblas.a}, \texttt{liblapack.a},
\texttt{libf2c.a}  (in \texttt{lib/i386/}).
The subdirectories with source code contain also the \texttt{Makefile}s so that
the executable files can be compiled with the \textsc{make} utility. By default the executable is put in the same directory.

\section{\label{s:test-calculations}Test calculations}
In order to facilitate a quick start, five examples for the application of the \textsc{Qprop} code are provided in the directories 
\texttt{src/hydrogen/}, \texttt{src/ionization/}, \texttt{src/neon/}, \texttt{src/winop/}, and \texttt{src/circ/}. The examples are chosen such that, on one hand  they are not too run-time consuming  while, on the other hand they cover in a non-trivial manner the key issues {\em (i)} imaginary time propagation, {\em (ii)} real time propagation, {\em (iii)} determination of an effective potential using DFT, {\em (iv)} calculation of electron spectra, and {\em (v)} circular polarization.

Each subdirectory in \texttt{src/} contains the source code in the files
with extension \texttt{.cc} and an appropriate \texttt{Makefile} to generate
the executable program in the same directory. The output will be written to files in the subdirectories {\tt res}.

\subsection{\label{ss:imag-time-prop}Ground state via imaginary time propagation}
The goal in this example is to calculate the nonrelativistic ground state of the hydrogen atom 
by means of imaginary time propagation.
As explained in Sec.~\ref{ss:imag-prop}, imaginary time 
propagation is a powerful method to compute the ground state in any potential. 
The hydrogen
atom was chosen because of its simplicity and the fact that the solution is analytically known.
We start with an unbiased guess, that is, a random $s$-orbital. 
The source code of the program \texttt{hydrogen\_im.cc} is located in the
subdirectory \texttt{src/hydrogen/}. In the following we discuss the few crucial points.

After standard directives concerning header files, the file
\texttt{potentials.cc} is included.
{\small
\begin{verbatim}
// *** potentials, nuclear_charge, and laser pulse parameters
//     are declared in potentials.cc ***
#include<potentials.cc>
\end{verbatim}}
The file \texttt{potentials.cc} is a piece of code, which defines a few global variables and the potentials that make up the Hamiltonian. The idea is to put the code common to imaginary and real time propagation in an extra file, thus reducing liability to inconsistencies. The laser field is off ({\tt E\_0=0.0}) during imaginary time propagation. 

Next, variables are declared and initialized. Let us consider
the lines below the initialization of the object \texttt{g} of class \texttt{grid}:
{\small
\begin{verbatim}   
// *** declare the grid ***
g.set_dim(34); // 44 elliptical polariz., 34 linear polariz.
g.set_ngps(1000,1,1);  // N_r, L, N
g.set_delt(0.2/nuclear_charge);  // delta r
g.set_offs(0,0,0);     // there is no offset in r, l, and N 
\end{verbatim}}
Depending on the polarization of the laser light, the suitable 
expansion of the wavefunction in spherical harmonics is chosen, i.e., either (\ref{general-expansion}) with all
$L^2$ radial orbitals, suitable for any polarization in the $xy$-plane, or (\ref{reduced-expansion}) for linear polarization along the $z$-axis (where $L$ radial orbitals suffice).
General and reduced expansion require different propagation procedures.
Thus a propagation mode has to be specified using the \texttt{set\_dim()} function.
The next line \texttt{g.set\_ngps(1000,1,1)} defines $N_r=1000$ spatial grid points for each radial orbital $\Phi_{ilm}(r t)$, $L=1$ $l$-quantum numbers (ranging from $0$ to $L-1$), and $N=1$ orbital. 
The hydrogen ground state
requires only a single orbital $\Phi_{1s}(r)$ of $s$-symmetry so that $L=1$ is chosen. Since {\tt nuclear\_charge} is 1, $\Delta r=0.2$ follows, and the equidistant grid reaches up to $r_{\max}=200$\,au. 
There are no grid offsets for the two propagation modes 34 and 44 discussed in this paper so that the grid initialization is completed by {\tt g.set\_offs(0,0,0)}. 

The Hamiltonian \texttt{hamilton} of 
class \texttt{hamop} is initialized through the functions defined in 
\texttt{potentials.cc}:
{\small
\begin{verbatim}
// the Hamiltonian
hamilton.init(g,vecpot_x,vecpot_y,vecpot_z,scalarpotx,scalarpoty,scalarpotz,
                imagpot,field);

// this is the linear and constant part of the Hamiltonian
staticpot.init(g.size()); 
staticpot.calculate_staticpot(g,hamilton);
\end{verbatim}}
\noindent The time-independent part of the Hamiltonian is stored in {\tt staticpot} (see {\tt cal\-culate\-\_staticpot} in Tab.~\ref{t:functions}).

The radial wavefunction is stored in the object \texttt{wf} of class
\texttt{wavefunction}. In this example there is only a single radial wavefunction because $NL=1$.  The lines below initialize and
normalize the wavefunction \texttt{wf} (dots ``{\tt ...}'' indicate lines of code that are omitted for the sake of clarity).
{\small
\begin{verbatim}  
// *** wavefunction initialization ***
wf.init(g.size()); 
wf.init(g,iinitmode,1.0,ells);
wf.normalize(g);
...

wf.dump_to_file_sh(g,file_wf_ini,1)
\end{verbatim}}
\noindent Depending on the argument \texttt{iinitmode} of type \texttt{int},
the wavefunction may be initialized either randomly or with
hydrogenic orbitals. The interested user may have a look at the corresponding {\tt init} function in {\tt wavefunction.cc}. In this example, the wavefunction is initialized
randomly ({\tt iinitmode = 1}) and written to {\tt file\_wf\_ini}.
(i.e., a file in the subdirectory \texttt{src/hydrogen/res/}).
The file consists of two columns of numbers (real and imaginary part) and $N_rLN$ lines, i.e., only a single radial orbital in this example.

The array {\tt ells[$i$]} is used to assign the $l$ quantum number $0$ to the radial wavefunction $\Phi_{l}$. This information is solely needed by the initialization routine {\tt init} when $L>1$ (see the more complex example \ref{ss:gs-ne} below).\footnote{If we choose $L=2$ and {\tt ells[0]=1} we obtain the $2p$ state since only the radial wavefunction $\Phi_{l=1}$ will be initialized with random numbers while $\Phi_{l=0}\equiv 0$. Note that there is no coupling between different $l$ quantum numbers during imaginary time propagation.}

The actual propagation is a loop: 
{\small
\begin{verbatim} 
  // *** imaginary time propagation ***
  for (ts=0; ts<lno_of_ts; ts++)
  {
    time = time + imag(timestep);  
    ...

    E_tot = real(wf.energy(0.0,g,hamilton,me,staticpot,nuclear_charge));
    ...

    wf.propagate(timestep, 0.0, g, hamilton, me, staticpot, 0, nuclear_charge);

    wf.normalize(g);
    ...

  }
\end{verbatim}
}
\noindent Being inside the body of the loop, 
the short-time propagator is applied consecutively
to the orbitals in \texttt{wf} with an imaginary  time step  {\tt g.delt\_x()/4.0}, i.e., $0.05$ here.  The short-time propagation
is handled by the \texttt{wavefunction} member function \texttt{propagate()} (cf.\ Tab.~\ref{t:functions}). Its first argument \texttt{timestep} is the pure imaginary time step. The second argument, i.e., the time, is irrelevant for the determination of the ground state and thus set to zero. Grid, Hamiltonian, process ID {\tt me}, constant potential part {\tt staticpot}, magnetic quantum number $m=0$, and {\tt nuclear\_charge} (as defined in {\tt potentials.cc})  follow as arguments.

In the loop body the total energy is calculated each imaginary time step so that one 
may keep track of the convergence of the
total energy. After each propagation step the function is normalized 
because propagation in imaginary time is not unitary.

Imaginary propagation stops when a key is hit, i.e., when the user is satisfied with the convergence of the ground state energy, or at latest after {\tt lno\_of\_ts = 640000} time steps.  After propagation, the orbital is stored in {\tt file\_wf\_fin} in \texttt{src/hydrogen/res/}. The converged ground state energy on the numerical grid chosen above reads $\energy=-0.5001510772159702$\,au.

The names of the files generated in \texttt{src/hydrogen/res/} depend on the simulation parameters and read in the case discussed here
{\small
\begin{verbatim} 
hydrogen_im-34-1000-2.00e-01.log
hydrogen_im_info-34-1000-2.00e-01.dat
hydrogen_im_obser-34-1000-2.00e-01.dat
hydrogen_im_wf_fin-34-1000-2.00e-01.dat
hydrogen_im_wf_ini-34-1000-2.00e-01.dat
\end{verbatim}
}
\noindent The first file is a log-file containing all the important simulation parameters for inspection by the user. 
The second file is a xml-file from which other programs may conveniently read the simulation parameters. Observables are stored in the third file (time step and energy in two columns). Fourth and fifth file contain the wavefunction after and before the imaginary time propagation.

\begin{figure}[htb]
\begin{center}
\epsfig{file=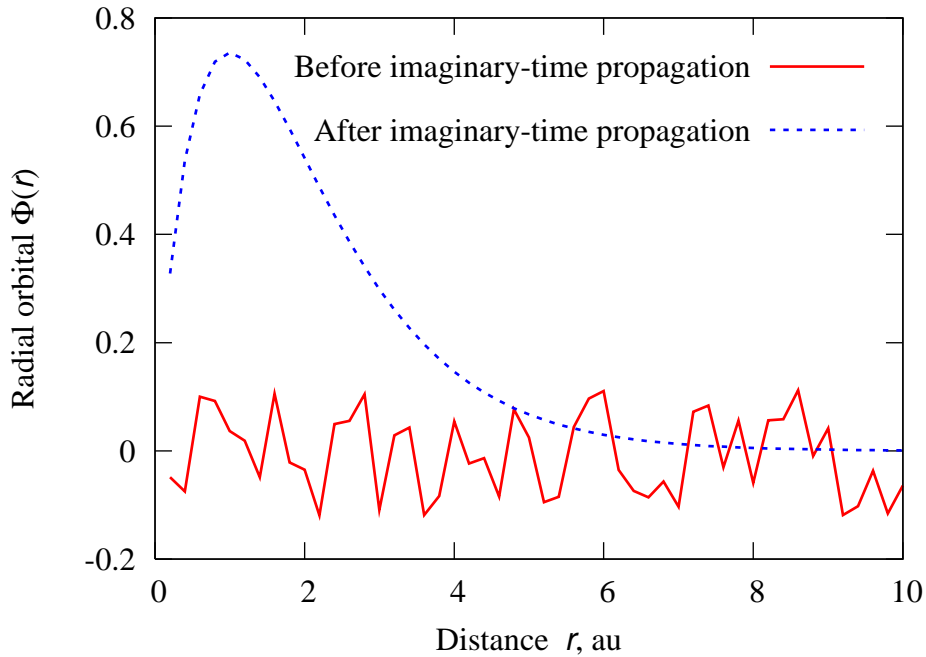, width=9.0cm, angle=270}
\end{center}
\caption{Initial (random) orbital before imaginary time 
propagation and final orbital converged to the hydrogenic
$1s$ radial wavefunction $2r\exp(-r)$.}
\label{f:example-1}
\end{figure}
The radial wavefunctions before and after imaginary time propagation are shown in Fig.\,\ref{f:example-1}. The wavefunction after imaginary time propagation converged to the $1s$ radial wavefunction $2r\exp(-r)$.

The program {\tt hydrogen\_re.cc} in the same directory reads the wavefunction from file {\tt hydrogen\_im\_wf\_fin-34-1000-2.00e-01.dat} and propagates it in real time with the laser field switched on. The essential lines are provided with comments so that the code should be self-explanatory. Additional entities of interest such as the time-resolved ground state population, the expectation value in laser polarization direction $\langle z\rangle$, and the total norm on the grid are calculated inside the main propagation loop.

\subsection{\label{ss:real-time-prop-h}One, two, and three-photon ionization of hydrogen}
The results presented in this example were obtained using the code in {\tt src/ionization}. First, the hydrogen $1s$ ground state is generated again using {\tt hydrogen\_im.cc}. Afterwards, the wavefunction is propagated in a $N_\mathrm{c}=20$-cycle $\sin^2$-pulse ($\sin^2$ with respect to the electric field) 
\beq \bE(t)=E_0\, \be_z\, \sin^2\left( \frac{\omega t}{2 N_\mathrm{c}} \right)\, \cos(\omega t+\varphi), \quad 0\leq t\leq T=N_\mathrm{c}\frac{2\pi}{\omega} \label{laserfield} \eeq
for different laser intensities $I$ between $10^{11}$ and $10^{12}$\,Wcm$^{-2}$ using {\tt ionization.cc}. The vector potential $\bA(t)$ corresponding to \reff{laserfield}, i.e., $\bA(t)=-\int_0^t \bE(t')\,\diff t'$,  is implemented in {\tt potentials.cc}. We want to focus on one and three-photon ionization. To that end we chose $\omega=0.8$ and $\omega=0.17$, respectively.

It is well known that if the ponderomotive energy 
\beq U_\mathrm{p}=\frac{E_0^2}{4\omega^2}, \eeq 
i.e., the time-averaged quiver energy of a free electron in a laser pulse of amplitude $E_0$, is much smaller than the ionization potential $I_\mathrm{p}>\hbar\omega$, ionization will be a perturbative process
in the multiphoton regime, and perturbation theory in lowest nonvanishing order (LOPT) 
is sufficient. 

The number of photons needed to overcome
the ionization potential of the atom is given by
\beq k  =  \left\lfloor \frac{I_\mathrm{p} }{ \hbar \omega} \right\rfloor\eeq
where $\hbar \omega$ is the energy of a photon and 
$\lfloor A\rfloor$ denotes the smallest integer $> A$. 
Thanks to the small ponderomotive energy the ac Stark shift can be omitted.

\begin{figure}[htb]
\begin{center}
\epsfig{file=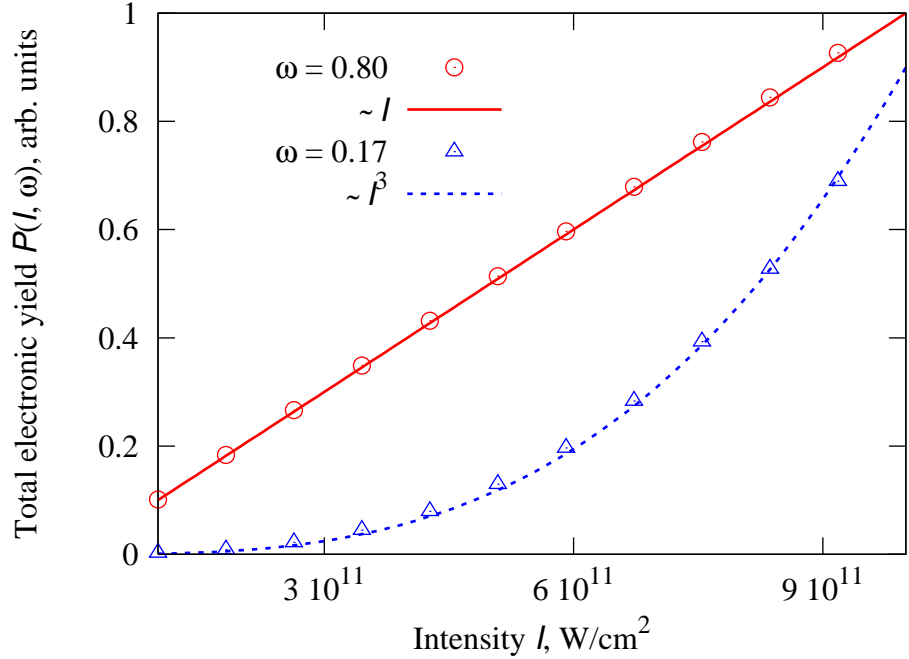, width=9.0cm, angle=270}
\end{center}
\caption{Ionization probability after a 20-cycle $\sin^2$-pulse vs laser intensity for the two carrier frequencies $\omega = 0.8$ and $\omega = 0.17$
 (one and three-photon ionization, respectively).
The power-law in \reff{power-low-mpi} is confirmed.}
\label{f:ex2}
\end{figure}

If the photon energy is smaller than the ionization potential,
one-photon ionization is impossible because of energy
conservation. On the other hand, $k+1$, $k+2$, $\ldots$ photon  ionization
is possible but unlikely in the LOPT regime. The latter  
predicts that the ionization rate $\Gamma_k(I,\omega)$
for a $k$-photon process is proportional to the $k$-th
power of the light intensity  (see, e.g., \cite{Delone-Krainov:1999}), i.e.,
\begin{equation}
\Gamma_k(I,   \omega)  = \sigma_k(\omega)   I^k,
\label{power-low-mpi}
\end{equation}
where the coefficient $\sigma_k(\omega)$ denotes the generalized cross section
for $k$-photon ionization. 
As long as $\Gamma_k T \ll 1$ the ionization probability is also proportional to $\Gamma_k$. Figure~\ref{f:ex2} shows the numerical results obtained with {\tt ionization.cc} in {\tt src/ionization}, confirming the power-law behavior in the multiphoton LOPT regime.

\subsection{\label{ss:gs-ne}Calculation of the neon ground state configuration}
In this example, the total energy of the neon atom will be calculated 
within the Kohn-Sham formalism of density-functional theory.
The theoretical background is presented above in Sec.~\ref{s:theory}.

In our nonrelativistic treatment, 
the total ground state energy of the neon atom can be calculated using 
only the three Kohn-Sham orbitals $\Psi_{1s}(\br)$, $\Psi_{2s}(\br)$, and 
$\Psi_{2p}(\br)$, occupied by $2$, $2$, and $6$ ``electrons'', respectively,
{\small
\begin{verbatim}    
// *** declare the grid ***
nuclear_charge = 10.0;
g.set_dim(34); // only 34 (linear polariz.) works in Kohn-Sham mode
g.set_ngps(10000,2,3);  // we need 2 angular momenta and 3 orbitals here
g.set_delt(0.01);
g.set_offs(0,0,0);
\end{verbatim}
}

The lines
{\small
\begin{verbatim}    
  really_propagate[0] = 1; // orbital 0 is to be propagated (not frozen!)
  really_propagate[1] = 1; // orbital 1 is to be propagated (not frozen!)
  really_propagate[2] = 1; // orbital 2 is to be propagated (not frozen!)

  degener[0]          = 2; // two electrons in 1s shell
  degener[1]          = 2; // two electrons in 2s shell
  degener[2]          = 6; // six electrons in 2p subshell

  ms[0]               = 0; // m quantum number of orbital 0 
  ms[1]               = 0; // m quantum number of orbital 1 
  ms[2]               = 0; // whether one puts 0,1, or -1 here 
                           // does not matter for the ground state 

  ells[0]             = 0; // l quantum number of orbital 0 
  ells[1]             = 0; // l quantum number of orbital 1 
  ells[2]             = 1; // l quantum number of orbital 2 
\end{verbatim}
}
\noindent are self explanatory. They assign $l$ and $m$ quantum numbers, degeneracies (i.e., occupation numbers), and the information whether orbital no.\ $i$ should be really propagated  ({\tt  really\_propagate[$i$]=1}) or considered frozen ({\tt  really\_propagate[$i$]=0}).

As compared to the simpler case of atomic hydrogen in Sec.~\ref{ss:imag-time-prop} the overloaded Kohn-Sham version of \texttt{propagate()} requires the radial functions $V_\mathrm{ee}^0$, $V_\mathrm{ee}^1$, and $V_\mathrm{ee}^2$ from the multipole expansion \reff{v_ee-expansion} as arguments:
{\small
\begin{verbatim}    
Lambdavector = wf.calculate_Lambda(g,degener);
U_H0 = wf.calculate_hartree_zero(g,Lambdavector);
U_x0 = wf.calculate_kli_zero(g,Lambdavector,ells,ms,
                             degener,islateronly,0);
V_ee_0 = U_H0 + U_x0;


wf.propagate(timestep, 0.0, g, hamilton, me, staticpot,
             V_ee_0, V_ee_1, V_ee_2, ms, nuclear_charge, really_propagate);
    
wf.normalize(g,ms);
\end{verbatim}}

Because all shells
in ground state neon are closed, the electron density distribution is spherical, and so is the effective potential.
Thus the monopole term of the effective potential
$V_\mathrm{ee}^{0}(r) = U^{0}(r)   +  V_{x}^{\mathrm{KLI}\  0}(r)$ is
sufficient. The same holds for open-shell systems in the commonly applied central field approximation.  The function {\tt normalize(g,ms)} performs a Gram-Schmidt orthonormalization of the Kohn-Sham orbitals. 
Exchange potential (in KLI approximation) and Hartree potential are calculated  using \texttt{calculate\_kli\_zero()} and \texttt{calculate\_hartree\_zero()}, respectively.

Next, a few characteristic energies are calculated and written to the file {\tt  file\_obser\_imag} 
{\small
\begin{verbatim}    
//
// calculate energies
// 
orb_energs = wf.orbital_energies(0.0, g, hamilton,me,
                                 staticpot, nuclear_charge);
orb_hartreesandexchange = wf.orbital_hartrees(0.0,g,me,U_H0+U_x0);
E_sp = wf.totalenergy_single_part(g,orb_energs, degener);
hartree_energy = wf.totalenergy_hartree(g,Lambdavector,U_H0);
x_energy=wf.totalenergy_exact_x(g,ells,degener);
E_tot = real(E_sp + hartree_energy + x_energy);
//
// output to observables file
//    
fprintf(file_obser_imag,"%i %15.10le %15.10le %15.10le %15.10le ",
        ts,E_tot,E_sp,real(hartree_energy),real(x_energy));
for (i=0;i<g.ngps_z();i++)  // loop over KS orbitals 
               fprintf(file_obser_imag,
               "%15.10le ",real(orb_energs[i]+orb_hartreesandexchange[i]));
\end{verbatim}}
\noindent  Here, we defined the orbital energy of Kohn-Sham orbital $i$ as
\beq  E_{\mathrm{orb}  i}=\left\langle \Psi_{i}\left|-\frac{\nabla^2}{2} - 
\frac{Z}{r}\right|\Psi_{i}\right\rangle. \label{sp}\eeq
The result is stored in the array {\tt orb\_energs}. The single-particle energy is defined as
\beq  E_{\mathrm{sp}} = \sum_{i} d_i  E_{\mathrm{orb}  i}, \eeq
i.e., the orbital energies $E_{\mathrm{orb}  i}$, weighted by the occupation numbers $d_i$. The total energy is then given by
\begin{equation}
E_{\mathrm{tot}} = E_{\mathrm{sp}} + E_{\mathrm{H}} + E_\mathrm{x}
\end{equation}
with $E_{\mathrm{H}}$ according \reff{H-pot} and $E_\mathrm{x}$ according \reff{x-energ}.
The orbital energies $\epsilon_i$ in \reff{kse} are given by
\beq \epsilon_i=E_{\mathrm{orb}  i} + E_{\mathrm{ee}  i} \eeq
where 
\beq  E_{\mathrm{ee}  i} = \int \diff^3 r\, \vert \Psi_i(\br)\vert^2 V_\mathrm{ee}(\br) = \langle V_\mathrm{ee}\rangle_i  \eeq
denotes the orbital contribution due to the electron-electron interaction.
In the code above, $ E_{\mathrm{ee}  i}$ is stored in the array  {\tt orb\_hartreesandexchange}. The sum of {\tt orb\_energs[$i$]} and {\tt orb\_hartreesandexchange[$i$]} gives $\epsilon_i$, which is written to file {\tt  file\_obser\_imag} as well. Note that owing to the nonlinearity of the Kohn-Sham equation $E_\mathrm{tot}\neq \sum_i d_i\epsilon_i$.

\begin{table}
\caption{\label{t:ne-energies}Orbital-, single-particle-, Hartree-, exchange-
and total energies for  the neon atom after imaginary time propagation.
The accuracy of the results improves with decreasing grid spacing $\Delta r=h$. The maximum radius  covered by the grid was fixed to $80$\,au in each calculation. Exact KLI values were taken from \cite{Grabo:2000,Ullrich-and-Gross:1997}.}
\begin{center}
\begin{tabular}{|c||cccc|p{1.7cm}p{2.0cm}|}
\hline
$h$, au $\rightarrow$ & $0.02$ & $0.01$ & $0.005$ & $0.0025$ & 
   exact KLI &
      LDA for Ne in \cite{NIST-DFTdata:2005}\\
\hline
$-\epsilon_{{1s}}$, au & $30.79983$  &$30.79928$&$30.80127$&
$30.80188$ &$30.80$&$30.3058$\\
$-\epsilon_{{2s}}$, au & $1.70687$  &$1.70711$&$1.70722$&
$1.70725$ &$1.707$&$1.32281$\\
$-\epsilon_{{2p}}$, au & $0.84941$ &$0.84941$&$0.84940$& 
$0.84940$&$0.8494$&$0.49803$\\
$-E_{\mathrm{sp}}$, au &  $182.6479$ &$182.6154$&$182.6137$ &
$182.6114$ &&$182.2495$\\
$E_{\mathrm{H}}[n]$, au& $66.20527$  &$66.17606$ & $66.16999$ &
$66.16588$ & &$65.72649$\\
$-E_\mathrm{x}$, au &            $12.11682$ &$12.10324$&$12.10013$ &
$12.09900$ & 12.099 &$11.71043$\\
$-E_{\mathrm{tot}}$, au&  $128.5595$ &$128.5426$&$128.5439$ &
$128.5446$& 128.5448 &$128.2334$\\
\hline
\end{tabular}
\end{center}
\end{table}

The numerical results for different grid spacings $\Delta r=h$ are collected in Tab.~\ref{t:ne-energies}. Very good agreement with the exact KLI values is obtained. Of particular interest is the orbital energy of the valence electron, i.e., $\epsilon_{2p}$. Imagine we want to use the effective potential $V_\mathrm{ee}$ for a single active electron calculation where only the valence electron is propagated while all others are kept ``frozen''. Such an approach will only lead to quantitative correct answers if the modulus of the orbital energy $\vert\epsilon_{2p}\vert$ is close to the first ionization potential (i.e., if Koopmans' theorem \cite{koopmans:1933} is satisfied). The ionization potential of neon is $I_\mathrm{p}=0.793$ showing that KLI yields an enormous improvement compared to the simpler local density approximation (LDA) which is $37$\% off.

A few remarks are advisable at that point. 
\begin{enumerate}
\item In the current version of \qprop\ the KLI potential is implemented for the search of the ground state configuration only. Unfortunately, the multipole expansion of the KLI expansion is utterly complicated if the Kohn-Sham orbitals loose their well-defined $l$ quantum number  (as it is the case when the laser field is switched on). For more details see the extra documentation in {\tt doc} or online  \cite{qpropwebpage}. The Hartree potential, on the other hand, can be evaluated using the time-dependent, perturbed Kohn-Sham orbitals up to the quadrupole. 
\item   As in any nontrivial optimization problem, it is not given for granted that the Gram-Schmidt orthonormalization 
during imaginary 
time propagation in the Kohn-Sham mode converges to the correct ground state configuration. The better are the initial guesses for the orbitals, the higher is the chance to find the true ground state. 
For the case of neon with only a few orbitals this is not a critical issue. We found out that for heavier atoms it is a good strategy to ``build up'' the electron configuration step by step, i.e., starting with an ion and adding one electron after the other. In that way we successfully generated the ground state configuration of xenon.
\item With a bad initial guess the orbital energies may cross during imaginary time propagation.
However, for a proper KLI potential it is important that the last orbital (i.e., the orbital with the highest index $i$) remains the outermost one during imaginary time propagation, since it is this orbital that is excluded from the sum in (\ref{KLI-eq}). Moreover, the sequence of orbitals must remain consistent with their degeneracies.
\item It is instructive (and a good test) to choose instead of the three orbitals above five orbitals $1s$, $2s$, $2p_{-1}$,  $2p_{0}$, $2p_{1}$, all occupied by $2$ electrons. The results in Tab.~\ref{t:ne-energies} are not affected, of course.
\end{enumerate}

\subsection{\label{ss:winop-example}Energy analysis of the final wavefunction}
Besides the actual time propagation, the \textsc{Qprop} package provides a tool to evaluate energy spectra and angular distributions
of the photoelectrons.
The spectral analysis of the final wavefunction is performed using the window-operator approach introduced in Sec.~\ref{ss:winop}. Here, we present an implementation of it called \textsc{Winop}.

The program {\tt winop.cc} is located in the subdirectory \texttt{src/winop}. It computes the probability distributions (\ref{pgne}) and (\ref{pgneo}) from a 
one-electron wavefunction $\Psi_i(\br)$ given in form of either the
expansion (\ref{general-expansion}) or expansion (\ref{reduced-expansion}).

First, the energy range and spectral resolution is defined: 
{\small
\begin{verbatim}
//
//  set a few essential parameters for the calculation of the spectra
//
lnoofwfoutput       = 1;      // 0 -- initial state, 1 -- final state.
no_energ            = 2750;   // number of energy bins
gamma               = 1.0e-4; // half width of the window operator (72)
energy_init         = -0.55;  // first energy; last energy then is
                              // energy_init + (no_energ-1)*2*gamma

// define for how many angles (77) should be evaluated 
no_theta            = 1; // no_theta angles between 0 and (1-1/no_theta)*pi   
no_phi              = 1; // no_phi angles between 0 and (1-1/no_phi)*2*pi 

\end{verbatim}
}
The variable {\tt lnoofwfoutput} specifies which of the stored wavefunctions should be considered for the calculation of the spectrum. This is important if several wavefunctions (or orbitals) are saved in the same file. In {\tt src/hydrogen/hydrogen\_re.cc}, for instance, the wavefunction is saved two times: before the propagation loop and after so that {\tt lnoofwfoutput} can be 0 (for an analysis of the initial wavefunction) or 1 (for the analysis of the final wavefunction). 

The next three parameters define the spectral range and resolution. The latter is governed by the parameter $\gamma$ of the window-operator \reff{window}: the smaller it is the higher is the spectral resolution. The variables {\tt no\_theta} and {\tt no\_phi} specify the number of angles for which the directional spectra   (\ref{pgneo}) are evaluated.

In order to ensure that the spectra are really calculated with respect to the proper Hamiltonian, the same {\tt potentials.cc}-file should be used for both the wavefunction propagation and the spectral analysis. For the latter only the unperturbed, spherically symmetric Hamiltonian is taken into account.

The grid associated with the wavefunction to be loaded is determined from the info-file generated by the propagation program (e.g., {\tt hydrogen\_re.cc} in {\tt src/hydrogen}) and stored in {\tt g\_load}. It is sometimes advisable to calculate the spectra on a grid that is larger than the grid on which the wavefunction was obtained because the energy resolution in the continuum increases when the grid radius is increased. In fact, our numerical grid is a spherical box whose discrete energy levels lie the closer together the bigger the radius is chosen. Since the level spacing increases with increasing energy, high resolution at high energies is only obtained for sufficiently large grid radii. Grid {\tt g} is used for the analysis. The loaded wavefunction is then regridded from {\tt g\_load} to {\tt g}.

The core of {\tt winop.cc} is a loop over the energy bins centered at $\mathcal{E}$.
Inside the loop, the function
{\small
\begin{verbatim}
 winop_fullchi( fullchi, result_lsub, &result_tot, energ, gamma,
                staticpot, V_ee_0, nuclear_charge, g, wf, iv);
\end{verbatim}
}
\noindent calculates the state $|\chi_{\mathcal{E}\gamma n}\rangle$ for order $n=3$ and gives it back as an object of class {\tt wavefunction}, named {\tt fullchi}. The latter is needed to evaluate   (\ref{pgne}) and (\ref{pgneo}).    The partial results 
\beq P_{lm \gamma n}(\mathcal{E})=\int \diff r\, R_{lm}^{\chi *}(r) R_{lm}^{\chi}(r) \label{partialspectra}\eeq
are of interest because they reflect selection rules related to angular momentum.  They are returned via the array {\tt result\_lsub}. Moreover, the partial spectra \reff{partialspectra} may be used to check the convergence of the wavefunction propagation with respect to the number of angular momenta $L$: if the partial contribution for $l=L-1$ is sizeable, $L$ should be increased. 

The output of {\tt winop.cc} to file {\tt file\_res} in the case of the restricted expansion (linear polarization) \reff{reduced-expansion} is organized as follows:

{\small
\bigskip
\noindent\begin{tabular}{|c|c|c|c|}
\hline
column 1 & columns 2 to $L+1$ & column $L+2$  & columns $L+3$ to $L+3+${\tt no\_phi*no\_theta} \\ \hline
Energy $\energy$& part.\  spectra Eq.~\reff{partialspectra} & total Eq.~\reff{pgne} & angular resolved Eq.~\reff{pgneo} \\ \hline
\end{tabular}
\bigskip
}

\noindent In the last {\tt no\_phi*no\_theta} columns the results are stored in the form 
\begin{eqnarray*} (\theta,\varphi)& =& (0,0), (\Delta\theta,0), (2\Delta\theta,0) \cdots (\pi-\Delta\theta,0), \\
 && \quad (0,\Delta\varphi),(\Delta\theta,\Delta\varphi),(2\Delta\theta,\Delta\varphi)\cdots (\pi-\Delta\theta,\Delta\varphi) \\
&& \qquad\qquad\qquad\vdots \\
 && \quad (0,2\pi-\Delta\varphi),(\Delta\theta,2\pi-\Delta\varphi) \cdots  (\pi-\Delta\theta,2\pi-\Delta\varphi)
\end{eqnarray*} 
with $\Delta\theta=\pi/${\tt no\_theta} and $\Delta\varphi=2\pi/${\tt no\_phi}.

In the case of the general expansion \reff{general-expansion} we have additional output for the various $m$ quantum numbers:

{\small
\bigskip
\noindent\begin{tabular}{|c|c|c|c|}
\hline
column 1 & columns 2 to $L^2+1$ & column $L^2+2$  & columns $L^2+3$ to $L^2+3+${\tt no\_phi*no\_theta} \\ \hline
Energy $\energy$ & part.\  spectra Eq.~\reff{partialspectra} & total Eq.~\reff{pgne} & angular resolved Eq.~\reff{pgneo} \\ \hline
\end{tabular}
\bigskip
}

\noindent  Columns 2 to $L^2+1$ contain the $P_{lm \gamma n}(\mathcal{E})$ in the form 
\[
(l,m) = (0,0),(1,-1),(1,0),(1,1),(2,-2),(2,-1) \cdots (L-1,L-1),
\]
i.e., $m$ is the faster running index.

The actual {\tt winop.cc} example-file the user finds in the {\tt src/winop}-directory reads the final wavefunction generated by the example for circular polarization in the following Sec.~\ref{ss:rlm-example}. However, the user may try to adapt the {\tt winop.cc}-file in order to analyze a final wavefunction generated by {\tt ionization.cc} described in
Sec.~\ref{ss:real-time-prop-h}.

\begin{figure}[htb]
\begin{center}
\epsfig{file=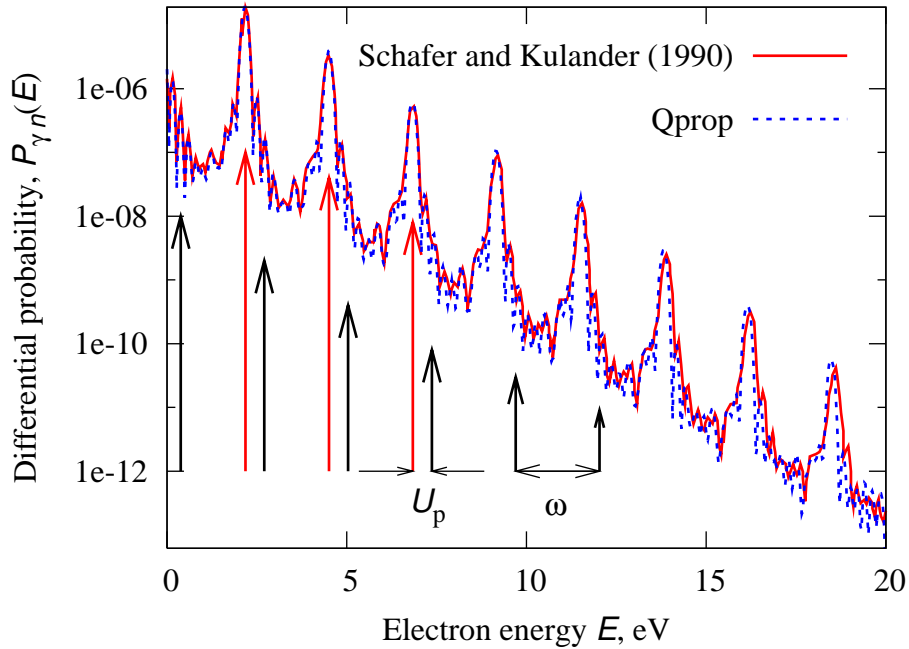, width=9.0cm, angle=270}
\end{center}
\caption{Continuum part of the electron spectrum of H($1s$) according Eq.~(\ref{pgne}) after the interaction with a 
linearly polarized, $\lambda = 535$\,nm laser pulse of intensity $2\cdot 10^{13}$ W/cm$^2$ with a trapezoidal profile in the electric field (up and down-ramped over 2 cycles, constant over 10 cycles). An energy bin of half-width $\gamma = 1.25 \cdot 10^{-2}$\,eV was chosen.
There is good agreement between the spectrum by Schafer and Kulander 
\cite{Schafer-and-Kulander:1990} and ours.
The formation of a sequence of peaks, all separated by $\hbar\omega$, is the manifestation of so-called  above-threshold ionization where the electron absorbs more photons than necessary to reach the continuum.
The shift of the peaks by the ponderomotive energy $U_\mathrm{p}$ is due to
the ac Stark shift of the continuum (with respect to the ground state).}
\label{f:example-4}
\end{figure}

As a test we reproduced the photoelectron spectrum
obtained by Schafer and Kulander in \cite{Schafer-and-Kulander:1990}.
The laser parameters were chosen as specified in \cite{Schafer-and-Kulander:1990}. They are explicitly given in the caption of Fig.\,\ref{f:example-4}. 
The photon energy $\hbar\omega=2.33$\,eV corresponds to six-photon
ionization in lowest order. Since we have to allow for multiphoton absorption with each photon increasing the angular quantum number $l$ by one,    $L$ was set to $15$. The fastest electrons to be resolved in the spectrum must not reach the absorbing boundary within the simulation time. To be safe, 40000 grid points in radial direction and $\Delta r=0.1$ were chosen. 

After having generated the final wavefunction with \qprop, \textsc{Winop} was used to calculate the spectrum shown in Fig.~\ref{f:example-4}. 
One clearly identifies eight above-threshold 
ionization peaks. The  peaks are located at 
$$\energy_k = -|\energy_{1s}| + k \omega - U_\mathrm{p}, \qquad k=k_{\min},\, k_{\min}+1,\, k_{\min}+2, \ldots $$
and separated by $\hbar\omega$. The $U_\mathrm{p}$-shift away from the ``naively'' expected position at  $-|\energy_{1s}| + k \omega$ is due to the ac Stark-effect. The integer $k_{\min}$ is the smallest integer $k$ that yields $\energy_k>0$.

The minor differences between the two curves in Fig.\,\ref{f:example-4} is due to the slightly different definition of $P_{\gamma n}(\mathcal{E})$ in \cite{Schafer-and-Kulander:1990}, the different orders $n$ used, and---most importantly---the final wavefunctions that were used for the analysis. The latter were obtained independently using different propagation algorithms.

\subsection{\label{ss:rlm-example}Rabi-flopping in a circularly polarized laser field}
In the last example we consider a hydrogen atom, driven resonantly by a circularly polarized laser field in dipole approximation,
\beq \bE(t)=\hat{E} [\cos(\omega t) \be_x - \sin(\omega t) \be_y]. \label{circpolfield}\eeq
The field vector rotates clockwise in the $xy$-plane.
The laser frequency is chosen to be resonant with the $1s \leftrightarrow 2p$-transition,
\[ \omega = \energy_{2p} - \energy_{1s} = \frac{3}{8}. \]  
Owing to the resonant driving a two-level approximation is adequate for not too high field amplitudes $\hat{E}$, i.e.,
\beq | \Psi(t) \rangle = d_{1s}(t) \exp(-\imagi \energy_{1s} t ) | 1s\rangle +  d_{2p}(t) \exp(-\imagi \energy_{2p} t ) | 2p_{-1}\rangle, \qquad  d_{1s}(0)=1, \ d_{2p}(0)=0 . \eeq
Here we assume that the atom starts in the $1s$ ground state, and we make use of the fact that the $2p$ level with $m=-1$ will be predominantly populated (as will be justified below). The time evolution of the two populations $|  d_{1s}(t) |^2$ and $|  d_{2p}(t) |^2$ is then governed by the differential equations
\begin{eqnarray} 
\frac{\diff}{\diff t}{d_{1s}}(t) &=& -\halb \imagi \hat{E}M_- \, d_{2p}(t), \\
\frac{\diff}{\diff t}{d_{2p}}(t) &=& -\halb \imagi \hat{E}M_-^* \, d_{1s}(t)
\end{eqnarray} 
where
\beq M_- = \langle 2p_{-1} | r \exp(i\varphi) \sin\theta | 1s\rangle = \frac{256}{243} . \eeq
As a consequence, the ground state population evolves according 
\beq  | {d_{1s}}(t) |^2 = \cos^2\left(\frac{\Omega_\mathrm{R}}{2}t\right) \label{rabigroundstatepop}\eeq
where $\Omega_\mathrm{R}$ is the so-called Rabi frequency (see any text book on quantum optics, e.g., \cite{scullyzubairy}). In our case one has
\beq \Omega_\mathrm{R}= M_- \hat{E} = \frac{256}{243} \, \hat{E}. \label{rabifrequ}\eeq

The example code can be found in the subdirectory {\tt src/circ}. Since we are dealing with circular polarization we need to employ the general expansion \reff{general-expansion} and therefore set {\tt g.set\_dim(44)}. The program {\tt circ\_im.cc} generates the hydrogen ground state on a numerical grid with $N_r=1000$, $L=1$, and $\Delta r=h=0.15$. The ground state wavefunction (stored in {\tt ./res/circ\_im\_wf\_fin-44-1000-1.50e-01.dat}) and the corresponding info file (stored in {\tt circ\_im\_info-44-1000-1.50e-01.dat}) are read by the program  {\tt circ\_re.cc}, which performs the actual time propagation. 

The propagation algorithm for elliptical polarization using the wavefunction expansion \reff{general-expansion} is significantly more involved than the method for linear polarization explained in Sec.~\ref{ss:propagation-algorithm}. A detailed description can be found online at \cite{qpropwebpage} and in the directory {\tt doc}. Here we only note that in propagation mode $44$ the vector potential must be of the form
\beq \bA(t)=A_x(t)\,\be_x + A_y(t)\, \be_y \eeq 
so that the Hamiltonian (with the purely time-dependent term $\sim \bA^2(t)$ transformed away) reads
\beq H(t)=-\frac{1}{2}\nabla^2+V(r)+V_I(t) \label{circHamilI}\eeq
where \beq V_I(t)= -\imagi A_x(t)\partial_x - \imagi A_y(t) \partial_y. \label{circHamilII}\eeq
The vector potential components $A_x$ and $A_y$ are defined in {\tt potentials.cc}. In order to match \reff{circpolfield} they are chosen as
\beq A_x(t)=-\frac{\hat{E}}{\omega} \sin(\omega t), \qquad  A_y(t)=-\frac{\hat{E}}{\omega} \cos(\omega t). \eeq

\begin{figure}[htb]
\begin{center}
\begin{tabular}{cc}
\epsfig{file=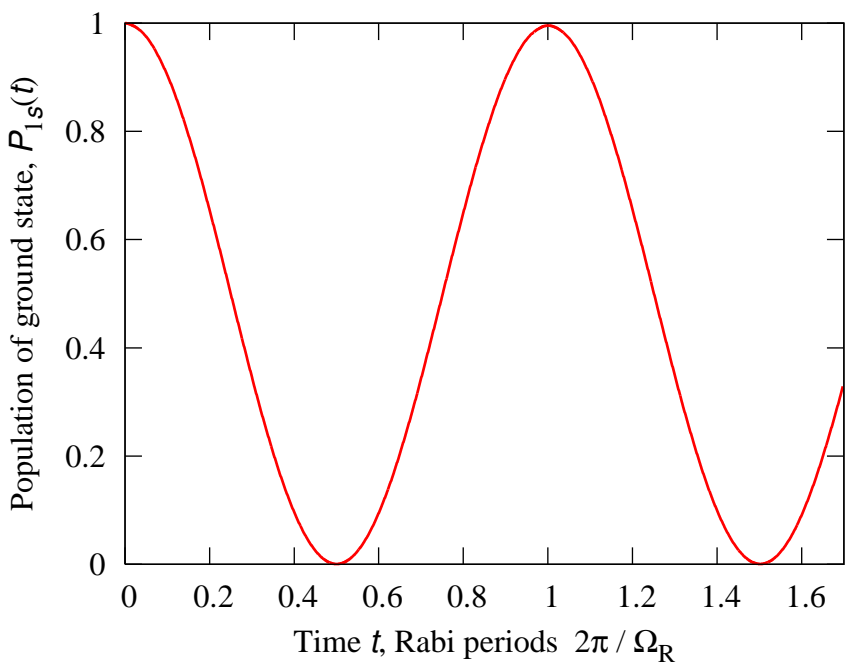, width=9.0cm, angle=270}
\end{tabular}
\end{center}
\caption{\label{rabiexamplefig1}Population $P_{1s}(t)=| d_{1s}(t)|^2$ vs time, rescaled with the help of the expected Rabi frequency \reff{rabifrequ} $\Omega_R=3.976\cdot 10 ^{-3}$.}
\end{figure}

Real time propagation is performed for $\hat{E}=3.774\cdot 10 ^{-3}$ (corresponding to $5\cdot 10^{11}$\,W/cm$^2$) on a numerical grid with $N_r=1000$, $h=0.15$, and $L=4$. The expected Rabi-frequency \reff{rabifrequ} is $\Omega_R=3.976\cdot 10 ^{-3}$. Fig.~\ref{rabiexamplefig1} shows that the temporal behavior of the ground state population (i.e., the third column in the output file {\tt ./res/circ\_re\_obser-0.375000-44-1000-1.50e-01.dat}) is in excellent agreement with Eq.~\reff{rabigroundstatepop}.

\begin{figure}[htb]
\begin{center}
\hspace{-3cm}
\begin{tabular}{p{0.41\linewidth}p{0.41\linewidth}}
\epsfig{file=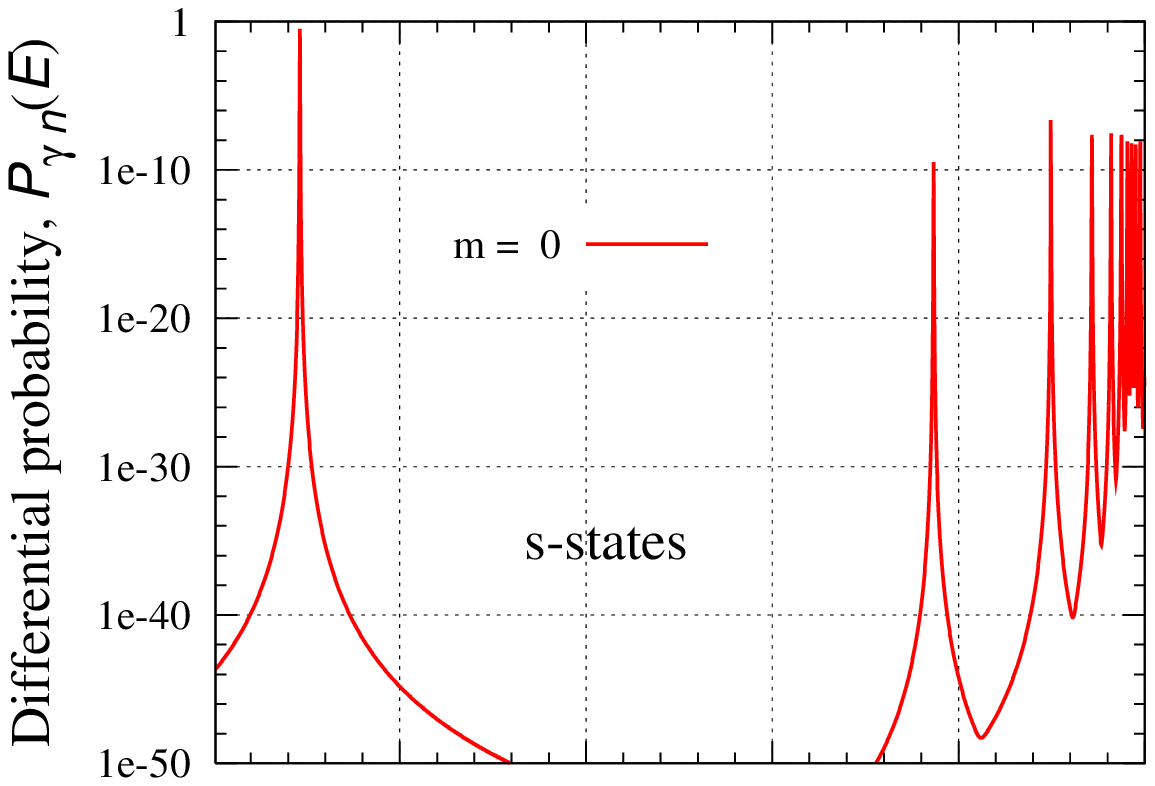, width=6.0cm, angle=270} &
\epsfig{file=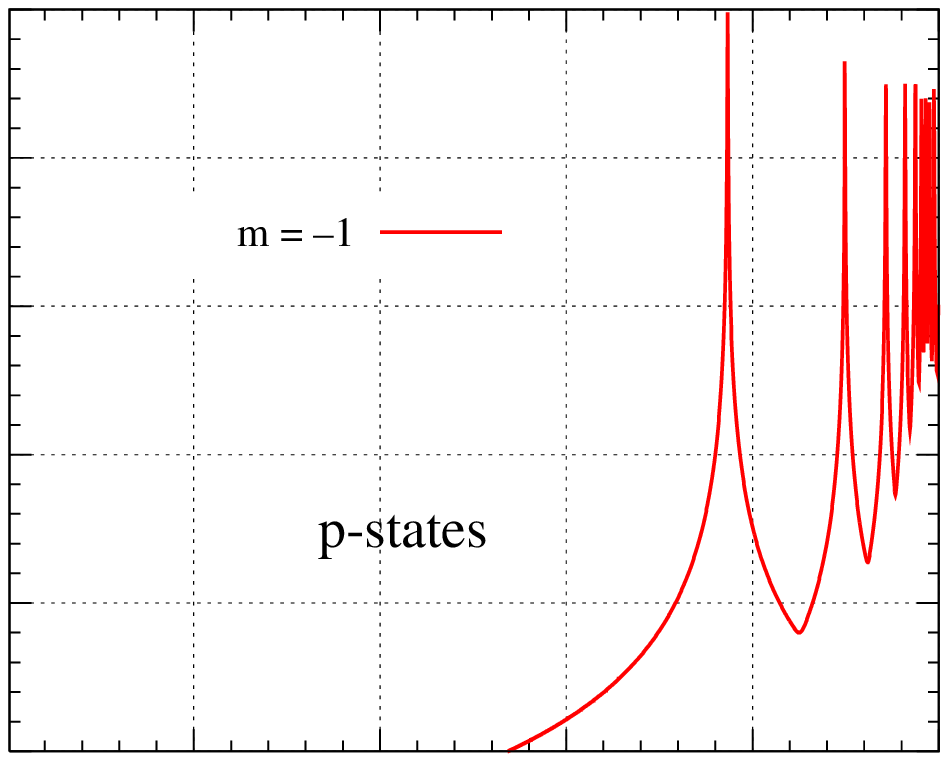, width=6.0cm, angle=270}
\\[-0.9cm]
\epsfig{file=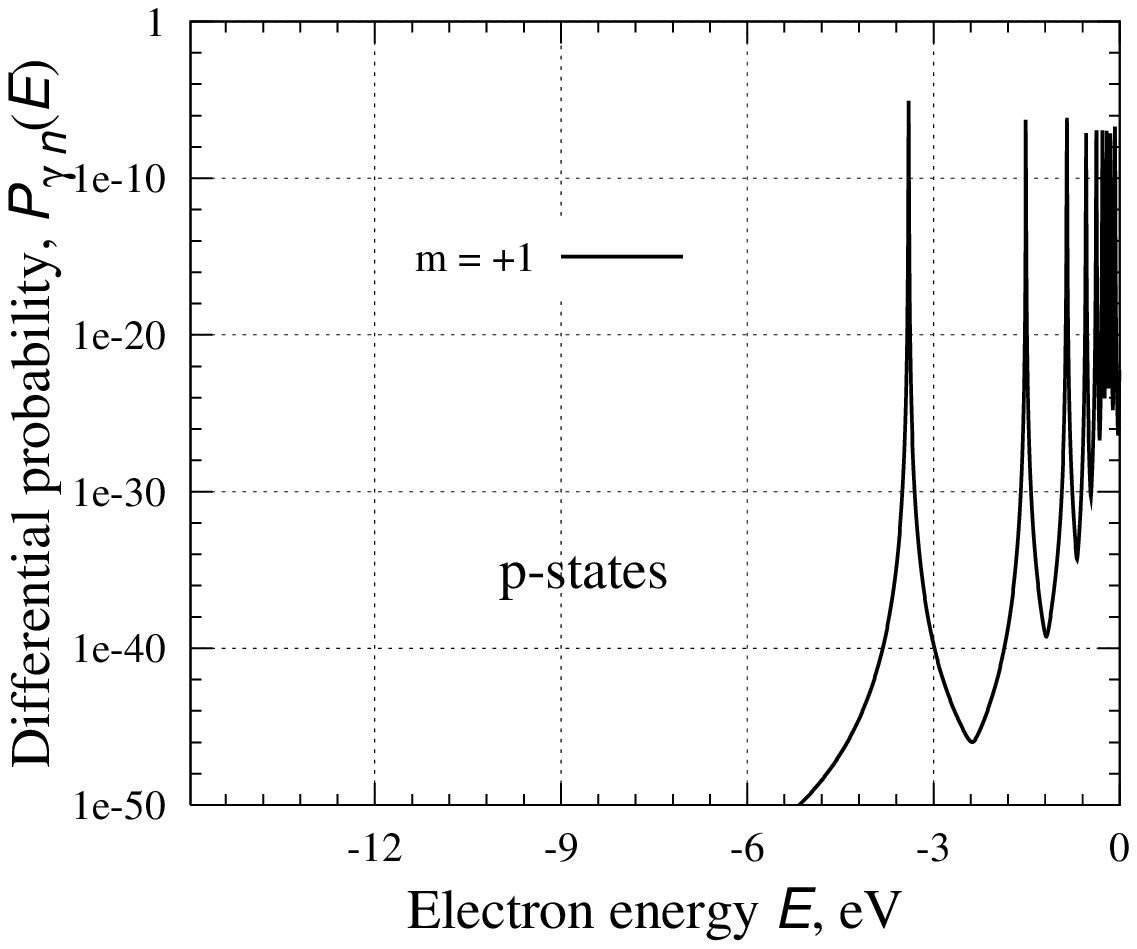, width=6.0cm, angle=270} &
\epsfig{file=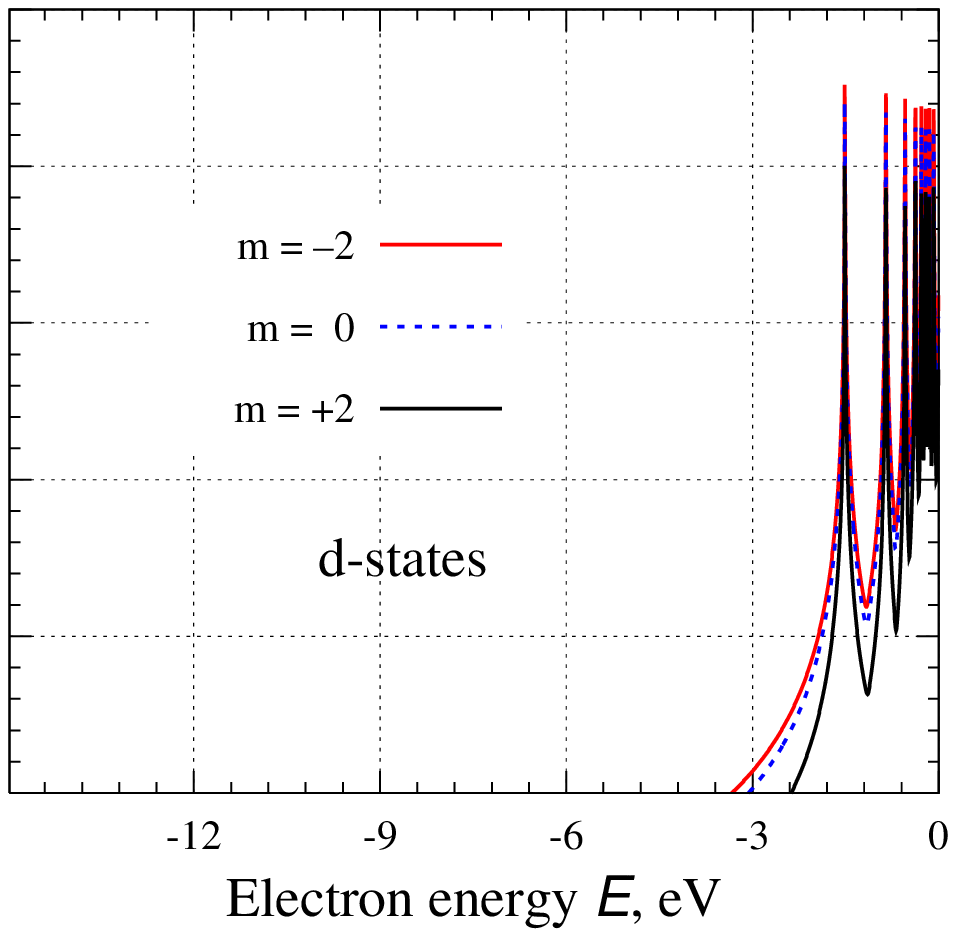, width=6.0cm, angle=270}
\end{tabular}
\end{center}
\bigskip
\caption{\label{rabiexamplefig2} Partial spectra for $l=0$ ($s$-states), $l=1$ ($p$-states), $l=2$ ($d$-states), $l=2$ ($f$-states), and the different $m$-quantum numbers (plotted in different colors). States with $m$s not shown are not populated at all. }
\end{figure}

Figure \ref{rabiexamplefig2} shows the result of a spectral analysis using the program {\tt winop\_circ.cc}\footnote{This file should be copied to {\tt winop.cc} before using {\tt make} for compilation.} in {\tt src/winop}, which reads the final wave function stored in {\tt src/circ/res/circ\_re\_wf-0.37\-5000-44-1000-1.50e-01.dat} and the corresponding info file. Moreover, {\tt potentials.cc} in {\tt src/circ} is included for the proper definition of the unperturbed Hamiltonian. The spectra are calculated in the energy range $[-0.55,0.0]$ with a resolution $\gamma=10^{-4}$.
On the logarithmic scale down to $10^{-50}$ and beyond in Fig.~\ref{rabiexamplefig2} many more states than just the $1s$ and the $2p_{-1}$ are populated. However, looking up the populations of the $n=2$-levels at $\energy=-1/8$ in the file {\tt ./res/res-0.375000-44-1000-4-2.000e-04-1.000e-04.dat} one infers that the next ``important'' state (the $2p_{1}$) is already a factor $\approx 10^{-5}$ less populated than the   $2p_{-1}$. Hence, the two-level approximation is very well applicable at  $5\cdot 10^{11}$\,W/cm$^2$. 

By virtue of the partial spectra in  {\tt ./res/res-0.375000-44-1000-4-2.000e-04-1.000e\--04.dat} it is also observed that the population of all states $p_0$, $d_{-1}$, $d_1$, $f_{-2}$,  $f_{0}$, $f_{2}$ remain {\em exactly} zero. This is because all couplings generated by the Hamiltonian \reff{circHamilI} (with a vector potential \reff{circHamilII}) require $\vert \Delta m \vert =1$ and $\vert \Delta l \vert =1$. As long as one starts with a well-defined state of the unperturbed system one could remove all the strictly unpopulated states in the expansion \reff{general-expansion}, thus saving a considerable amount of run-time. In cases where, for instance, the initial state is the ground state one could replace the general expansion \reff{general-expansion} by 
\[ \Psi_{i}(\br t)=\frac{1}{r}\,\sum_l \sum_{m+=2}  \Phi_{ilm}(rt)\,\rY_{lm}(\Omega), \]
where $\sum_{m+=2}$ runs in steps of $2$, i.e., $m=-l,-l+2, \ldots, l$.
However, we did not implement this in the present version of \qprop\ in order to allow also for the propagation of superpositions of states with different $m$s. 

It is interesting to increase the laser intensity so that more and more states---including the continuum---become involved, and the two-level approximation breaks down. Note that for obtaining converged results both $N_r$ and $L$ must be increased appropriately. It is known that at higher laser intensity the lowest-order photoelectron peak at $\energy\approx\energy_{1s} + 2\hbar\omega$ splits into two, separated by the Rabi frequency $\Omega_\mathrm{R}$ \cite{LaGattuta}.  The interested reader may try to reproduce this result using \textsc{Qprop}.

\subsection{Other tests}
Besides Kohn-Sham orbital energies and the comparison in Fig.~\ref{f:example-4}, additional tests of the \textsc{Qprop} package have been performed. Ionization rates were determined from the  time-dependent norm on the numerical grid and compared with the  highly precise ionization rates obtained with the help of the Floquet code \cite{potvliege}. Excellent agreement is obtained for sufficiently small $\Delta r$ and $\Delta t$, and sufficiently large grid. The grid size is either determined by the excursion $\hat{E}/\omega^2$ of the electrons in the laser field or the distance travelled by the fastest electrons that should be resolved in the final photoelectron spectra. If it is not necessary to resolve electrons beyond a certain threshold energy they may well be absorbed by the imaginary potential without spoiling the observables of interest. In any case, checking the results for convergence with respect to grid size, $\Delta r$, and $\Delta t$ is a must. 

The code was further tested by propagating a free Gaussian wave packet in a laser field---a problem that can be solved analytically. If, with increasing laser amplitude (i.e., increasing excursion) both the number of grid points in radial direction and the maximum $l$-quantum number $L-1$ were increased appropriately, excellent agreement was obtained. 

Our simulations of high order harmonic generation in many-electron atoms indicate that the truncation of the Hartree potential expansion after the quadrupole is safe: the spectra hardly change when the quadrupole term is omitted. Even the monopole alone is sufficient for a wide range of laser parameters.

\section{\label{s:summary-and-outlook}Summary and Outlook}
The \textsc{Qprop} package has been introduced. 
The purpose of the package is to study atoms or other (initially) spherical systems in strong laser fields. \textsc{Qprop} allows to investigate 
a great variety of nonperturbative phenomena including above-threshold ionization, high-order 
harmonic generation, and stabilization. The full time propagation of nonrelativistic, one-electron wavefunctions is
realized. Effective 
potentials incorporating electron-electron
repulsion and exchange can be taken into account.
Further development will be aimed at
removing the following restrictions:

The external field is currently treated in dipole approximation only, i.e., $\exp[\imagi(\omega t- \mathbf{k}\br)]$ is approximated by $\exp(\imagi\omega t)$. Going beyond the dipole approximation by taking into account the next order, 
$\exp[\imagi(\omega t-\mathbf{k}\br)] \approx \exp(\imagi\omega t)(1 - \imagi\mathbf{k}\cdot\br)$, allows to study new phenomena related to the $\mathbf{v}\times\mathbf{B}$-electron drift in $\mathbf{k}$-direction. The azimuthal symmetry is broken then so that (\ref{general-expansion}) has to be employed. 

Many-electron systems can be currently treated only on a 
Kohn-Sham level using either the local density approximation or KLI. 
Although \textit{ab initio} in principle, many 
important observables cannot be calculated in practice because their functional dependences on the Kohn-Sham orbitals are unknown. 
If, on the other hand, the full, generally correlated, many-electron wavefunction (or a reasonable approximation to it) is available, the computation of observables is straightforward. 
However, already the time propagation of a two-electron wavefunction is a demanding task \cite{Smyth-etal:1998,Meharg-etal:2005} as soon as the continuum becomes involved. 
For few-electron systems, the time-dependent multi-configurational Hartree-Fock method or related approaches are worth being investigated. Our current activities show that the implementation of such kind of schemes benefits heavily from the routines provided by \textsc{Qprop}.

%
\subsubsection*{Acknowledgement}
This work was supported by
the Deutsche Forschungsgemeinschaft.

{\renewcommand\baselinestretch{0.95}

}

\end{document}